\documentclass[aps,prl,reprint]{revtex4-1}
\usepackage{graphicx}
\usepackage{mathrsfs}
\usepackage{amsmath}
\usepackage{subfigure}
\usepackage{float}

\begin{document}

\title{The classical mechanics of autonomous microscopic engines}

\author{Lukas Gilz}
\author{Eike Thesing}
\author{James R. Anglin}

\affiliation{\mbox{State Research Center OPTIMAS and Fachbereich Physik,} \mbox{Technische Univerit\"at Kaiserslautern,} \mbox{D-67663 Kaiserslautern, Germany}}

\date{\today}

\maketitle

\onecolumngrid
Even microscopic engines have hitherto been defined to require macroscopic elements such as heat reservoirs, but here we observe that what makes engines useful is energy transfer across a large ratio of dynamical time scales (\textit{downconversion}), and that small, closed dynamical systems which could perform steady downconversion (``Hamiltonian daemons'') would fulfill the practical requirements of autonomous microscopic engines. We show that such daemons are possible, and obey mechanical constraints comparable to, but different from, the laws of thermodynamics.
\\
\twocolumngrid

Is it possible to miniaturize an entire engine so extremely that it consists of only a few degrees of freedom --- including its fuel supply? With advancing nanotechnology this question will ultimately be practical\cite{biomotor1}, but first it is conceptual, because engines have hitherto been defined to require macroscopic elements such as heat reservoirs\cite{Alicki,Schwabl, Callen,astumian_thermodynamics_1997,Kosloff,kieu_second_2004,hanggi_artificial_2009,blickle_realization_2011,kim_quantum_2011} or external control parameters with predetermined time-dependence\cite{Alicki,Kosloff,fialko_isolated_2012}. In the fundamental terms of mechanics, however, heat engines are simply systems in which a particular slow time evolution in one system (work) is achieved by transferring energy from another system whose time evolution is \textit{rapid} and \textit{chaotic} (heat)\cite{Clausius,Callen}. Statistical mechanics derives the constraints of thermodynamics from assumptions about the ergodic evolution associated with chaos, which can be absent in small systems; but it is not the chaos of heat that makes heat engines useful. It is the rapidity.

The dimensions of energy are Mass$\times$Length$^2$/Time$^2$, and so for given mass and size, energy depends on time scale. Heat engines are ultimately governed by the same laws of motion that govern clockwork, but engines can do more work than coiled springs, because fuel stores energy more densely --- because molecular dynamics is fast. A car engine turns wheels at a few thousand RPMs, using energy stored in molecular degrees of freedom with frequencies in the $10^{14}$ Hz range (electron-Volt energy scales). The power advantage of engines over clockwork is thus due to an enormous ratio of time scales.

A fully microscopic closed system with only a few relevant degrees of freedom, and no chaotic subsystem, would not obviously be limited by conventional statistical mechanics or thermodynamics; it would answer only to pure mechanics. If it could nonetheless steadily perform work using energy from high-frequency degrees of freedom (\textit{steady downconversion}), it could deliver power like a tiny heat engine, but without disorder. We refer to such a system as a \textit{daemon engine}, or for brevity \emph{daemon}, by analogy with the daemons of computer operating systems (autonomous background processes, originally so named as deterministic analogs of Maxwell's Demon). Physically realized daemon engines might one day let autonomous nanobots operate for extended periods under onboard power. Here we consider daemons theoretically, as a basic dynamical phenomenon. Sufficiently small daemon engines may require quantum mechanical description, but as a first step we restrict our attention to classical mechanics, which remains a valid approximation in many microscopic regimes. 

\begin{figure}[htbp]\label{fig:smallfigure05}
	\includegraphics[width=.4\textwidth]{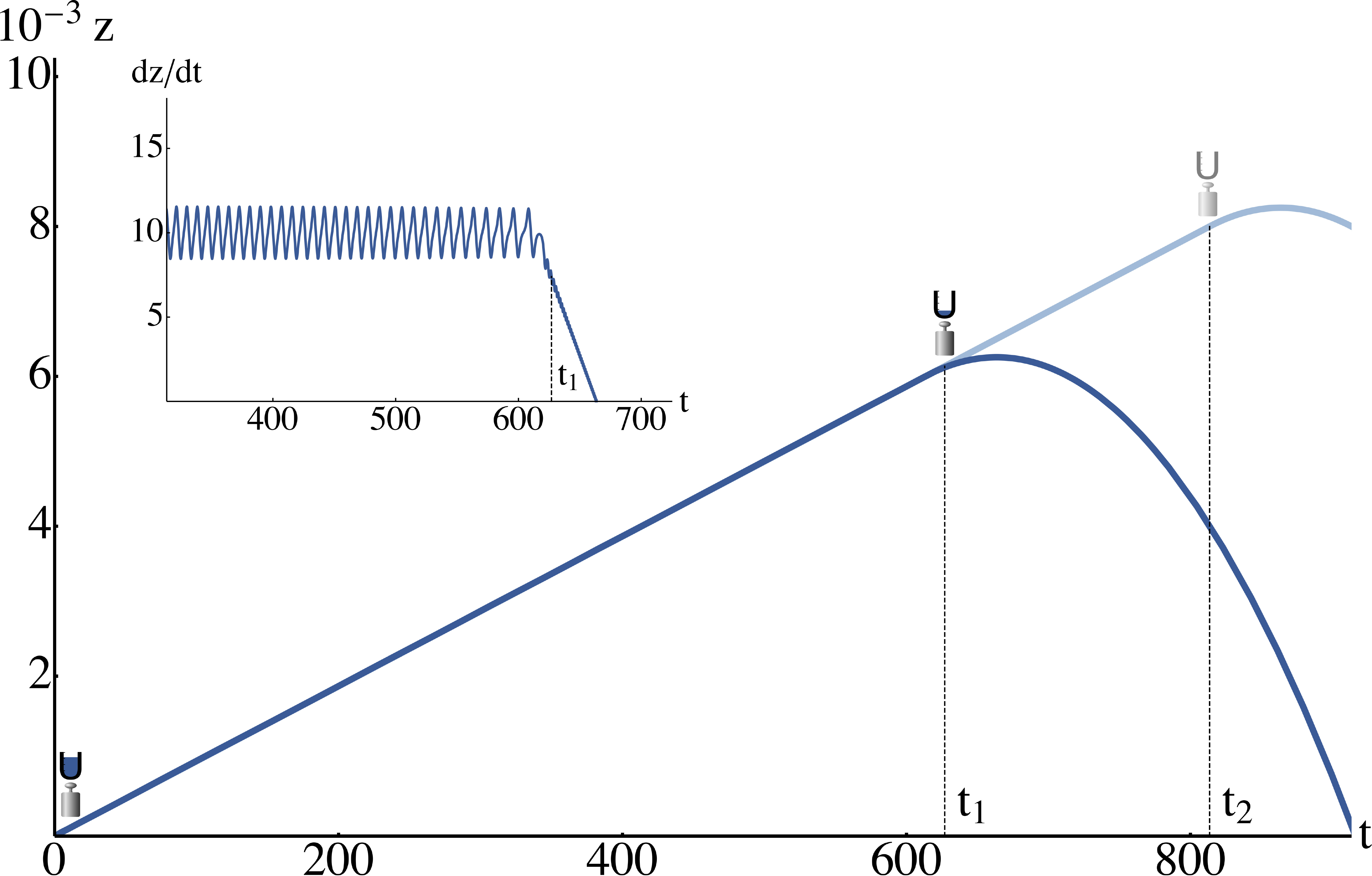}
	\caption{Example of daemon dynamics: in a closed system, a weight is steadily lifted using energy from a high-frequency subsystem. (Inset: the weight's velocity actually oscillates, as the daemon engine cycles, around a steady mean.) The filling level of the $U$-shaped fuel tank above the weight represents the energy of the high-frequency subsystem.  At time $t_{1}$ a dynamical transition occurs: the daemon engine stalls and the weight decelerates and falls under gravity. Stalling occurs at $t_{1}$ for reasons of phase space volume as explained in the text, even though approximately 25\% of initial energy remains in the high-frequency subsystem; energy conservation alone would have allowed the weight to rise higher (light curve). The plotted $z(t)$ is a numerical solution to the equations of motion under the Hamiltonian (\ref{EngineH}) with the particular coupling $V(z-\tau,U)=-\sqrt{U(2-U)}\cos(z-\tau)$, small parameters $\epsilon=0.01$, $\delta=0.1$, $g=0.2$, and initial conditions $p(0)=0.1$, $z(0)=0$, $U(0)=1.6$, $\tau(0)=0.9\pi$. }
\end{figure}

Classical mechanics determines motion in the abstract \textit{phase space} of generalized position and momentum coordinates $(q_n,p_n)$, under the universal equations
\begin{equation}\label{eq:Canon}
\frac{dq_n}{dt} = \frac{\partial H}{\partial p_n}\quad,\quad \frac{dp_n}{dt} = -\frac{\partial H}{\partial q_n}
\end{equation}
where the \textit{Hamiltonian} function $H(\{q,p\}_n)$ defines the system's total energy. Conservation of energy is a fundamental law because (\ref{eq:Canon}) implies $dH/dt = 0$, no matter what form $H$ has. If all relevant degrees of freedom are included in the model, any physical system can be represented by some $H$. We now consider what kind of $H$ might represent a daemon that can do work by steady downconversion within a closed system.

Any work task can be modeled as lifting a weight\cite{Carnot}, so a daemon system includes a weight which can be lifted at some steady speed, using energy from another subsystem (the `fuel') whose dynamics is rapid but not chaotic. Using Hamilton-Jacobi variables to describe the integrable fuel in terms of fuel energy $U$ and its fast conjugate variable $\tau$, and discarding small non-resonant terms whose effects will be negligible, we can express any daemon Hamiltonian as 
\begin{eqnarray}\label{EngineH}
H = \frac{{p}^2}{2\epsilon} + \frac{U}{\delta} + \epsilon\, g z + \epsilon V(z-\tau,U,\vec{\alpha})\;,
\end{eqnarray}
where $z$ and $p$ are the height and upward momentum of the weight to be lifted, $g$ is the external force against which work is done, and $\epsilon$ and $\delta$ are small parameters. We use $\vec{\alpha}$ to denote any additional degrees of freedom that may be present. The coupling $V$ is bounded and $U$ is bounded from below. See the Supplementary Material for a derivation of the form (\ref{EngineH}) as representative of all daemons that are small, can do a lot of work, and work steadily. The crux of that derivation is the dependence of $V$ on $z$ and $\tau$ only in the combination $z-\tau$, because no other dependences are resonant near the steady lifting speed $dz/dt = 1/\delta$. The lifting speed is large (in the dimensionless microscopic units) so that the daemon delivers the desired high power for its size.

Examining (\ref{eq:Canon}) for the $H$ of (\ref{EngineH}) reveals three time scales. The fastest evolution is that of $\tau$, with $\dot{\tau}=\delta^{-1}+\mathcal{O}(\epsilon)$ --- and since $\tau$ is a linear combination of (at most a few) \emph{periodic} angle variables, this corresponds to high-frequency motion. The slowest is that of $U$ and $\vec{\alpha}$, whose time derivatives are of order $\epsilon$. In between is the time scale of $z$, since $d^2z/dt^2 = \mathcal{O}(1)$. Given the wide separation of these three time scales, and the weak coupling of $U$ to $z$ through $\epsilon V$, the default expectation is adiabatic decoupling: one should be able to describe the weight's motion with an effective Hamiltonian from which the fast variable $\tau$ has been adiabatically eliminated, and find no steady transfer of energy from $U$ to the weight.

We now observe, however, that (\ref{EngineH}) provides $dJ/dt = 0$ for
\begin{eqnarray}\label{J}
J = U + p + \epsilon g t\;,
\end{eqnarray}
\textit{i.e.} $J$ is a constant of the motion (despite its explicit dependence on $t$). We therefore obtain an \emph{exact} effective Hamiltonian which describes the remaining lower-frequency dynamics in terms of canonical pairs $(q=z-\tau,\, p)$ and $(\tau,\, J)$
\begin{eqnarray}\label{Heff}
H_\mathrm{eff} &=& \frac{(p-\epsilon/\delta)^2}{2\epsilon} + \epsilon V_\mathrm{eff}(q,\epsilon t)\\
\label{Veff}
V_\mathrm{eff}&=& g\,q +V(q,\,J-p-\epsilon g t),\vec{\alpha}(\epsilon t))\;.
\end{eqnarray}
The explicit time dependence of $H_{\mathrm{eff}}$ is introduced by the canonical transformation. The original $H$ (\ref{EngineH}) remains time-independent and conserved.

We can now confirm that, throughout most of $(q,p)$ phase space, adiabatic decoupling is indeed the rule. Except where $\tilde{p}=p-\epsilon/\delta$ is small, the bounded potential $V$ is small compared to the large kinetic energy term in (\ref{Heff}), and its argument $z-\tau$ varies rapidly. In these circumstances, $V$ can be adiabatically neglected, and only the unbounded $gq$ term considered. This simply drives $p\sim -\epsilon gt$, and then the constancy of $J$ implies that $U$ is also constant: no downconversion occurs and the weight accelerates downward under gravity as usual. If $\tilde{p}$ is small, however, this means $z\sim t/\delta$, and the $(z-\tau)$ argument of $V$ will be slow. One could therefore expect a brief resonant transfer of energy between $U$ and $z$ as the weight's speed passes this critical value $\tilde{p} = 0$.

But what if $\tilde{p}$ remains near zero? Any local minima in $V_\mathrm{eff}$ will support bound orbits in which $\tilde{p}$ oscillates around 0. Since this means $p$ is merely oscillating around a constant value, $U$ must steadily sink to keep $J$ constant, while the weight steadily rises at $\dot{z}=p/\epsilon\sim 1/\delta$. The reconciliation of this behavior with familiar concepts of resonance is that the highly nonlinear daemon system has a velocity-specific resonance between $U$ and $z$, at the critical velocity $\dot{z}=1/\delta$, and so at this critical speed the weight actually can draw energy resonantly from $U$. Given only a local minimum in $V$, this power can then automatically maintain the critical weight speed against gravity. Steady downconversion can thus be self-sustaining within a closed Hamiltonian system.

Self-sustaining velocity-specific resonance is the basic dynamical mechanism which lets Hamiltonian daemons perform as autonomous microscopic engines. Yet it gives them features surprisingly reminiscent of macroscopic engines. The velocity locking that keeps a daemon working is dynamical: the small oscillations of $q$ and $p$, in bound orbits around a local minimum of $V_\mathrm{eff}$, correspond to small oscillations of the weight's upward speed, around the steady average of $1/\delta$.  The oscillating $(q,p)$ describe cycles in phase space, analogous to the cycles in pressure-volume space of macroscopic engines. Moreover, just as this cycling is slow compared to the fuel frequency, it is fast compared to the $\sim\epsilon$ rate at which fuel energy is drained. Insofar as $\epsilon$ may be arbitrarily small, the daemon cycle may be repeated indefinitely, in the same sense with which one says that a heat engine cycle can be repeated indefinitely, although in reality fuel is always finite.

Since fuel is finite, a working daemon must eventually stop working; this is enforced by the dependence of $H_\mathrm{eff}$ on $\epsilon t$. As we show in the Supplementary Material, any local well in $V_\mathrm{eff}$ must eventually start becoming shallower, because the boundedness of $U$ from below implies that $V_\mathrm{eff}$ can have no local minima in $q$ when $U=J-p-\epsilon g t$ is at its minimum value. The amplitude of $(q,p)$ oscillations in the slowly time-dependent $V$ well, however, remains finite, since their phase space orbits preserve their enclosed phase space area as an adiabatic invariant\cite{Goldstein}. At some point before all $U$ is converted into work on the weight, therefore, the $V$ well must become too small to support a bound orbit of the invariant area, and the effective $(q,p)$ particle must escape. Downconversion is replaced by adiabatic decoupling: the daemon engine has stalled, after converting some but not all high-frequency energy into work. The fact that daemons must generally leave some $U$ unconverted is a limitation similar but not identical to the thermodynamic limits on heat engine efficiency. And there is a further similarity yet.

Since $\epsilon$ is small, the bound orbits needed for steady downconversion occur only within a small region of $(q,p)$ phase space. A theorist can simply specify initial conditions within this region, but in fact initial state preparation is itself a physical time evolution, and an engine which requires external fine-tuning of initial conditions of its high-frequency variables is not autonomous in the fullest sense. Can daemons ever start by themselves, from generic initial conditions? Yes --- under non-trivial but achievable conditions.

$V_\mathrm{eff}$ must always depend on $\epsilon g t$, but it is only required to become shallower within some neighborhood of the minimum $U$. In higher ranges of $U$, $V_\mathrm{eff}$ may for some time rise, depending on the particular form of $V$.  It is then possible for a local maximum of $V_\mathrm{eff}$ to rise after $q$ has passed it, so as to trap $q$ like a door closed behind. See Fig.~2a. Conversely, a $V_\mathrm{eff}$ maximum which is sinking can never capture the system into a bound orbit. Since the explicitly time-dependent $V_\mathrm{eff}$ might also change the value of $H_\mathrm{eff}$, the rigorous statement of daemon ignition concerns not the height of the barrier, but the total phase space area of bound orbits. 

\begin{figure*}[ht]\label{fig:bigfigure}
	\subfigure[]{\includegraphics[width=.259\textwidth]{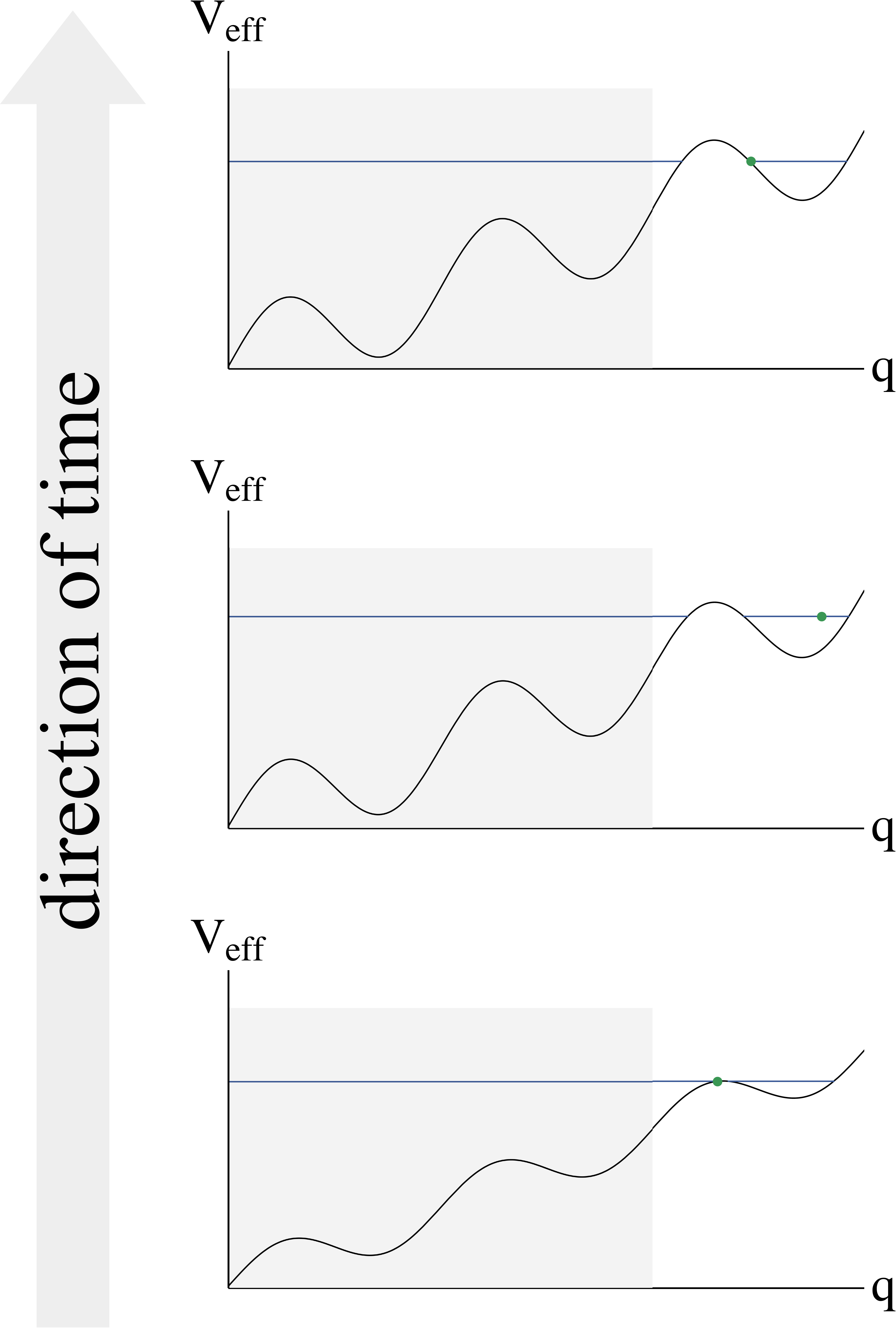}}
	\subfigure[]{\includegraphics[width=.345\textwidth]{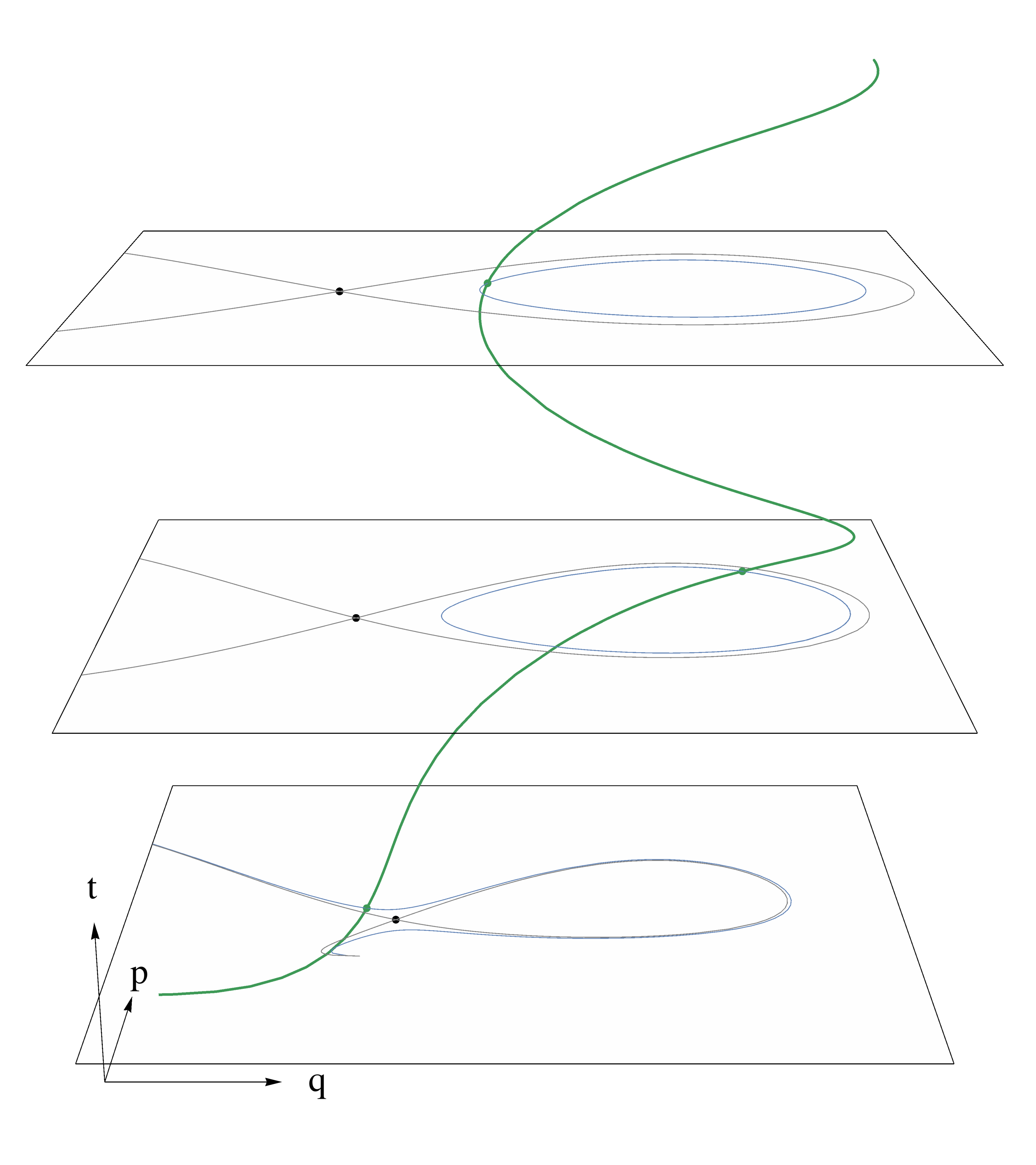}}
	\subfigure[]{\includegraphics[width=.345\textwidth]{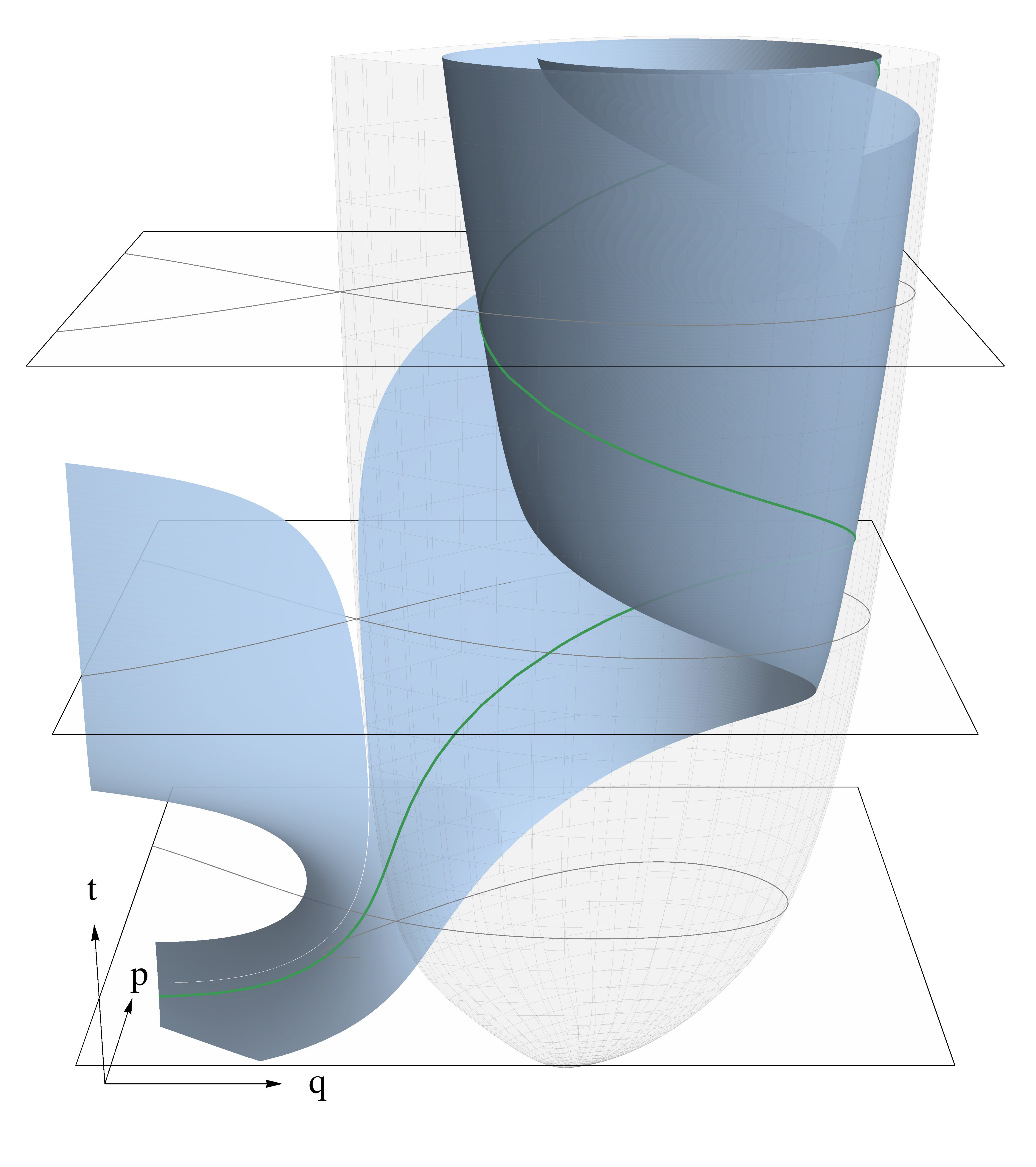}}
\caption{
Daemon autonomous ignition depicted as a) dynamical trapping in the effective potential $V_{\mathrm{eff}}(q)$; b) transition between unbound and bound orbits in phase space; c) evolution of a sheaf of trajectories in phase space.\\
\textbf{a)} Only the unshaded regions will be reproduced in b) and c) plots. Initially (bottom), the effective potential is shallow and the example trajectory (green dot) may pass the potential barrier (black line). Later (middle), $V_{\mathrm{eff}}$ rises and the system is trapped.\\
\textbf{b)} Black dot (unstable instantaneous fixed point) corresponds to the local maximum of $V_{\mathrm{eff}}$ in a). Initially (bottom), the example trajectory (green line) is adiabatically following an open contour of constant $H_{\mathrm{eff}}$ (blue contour), outside the separatrix (black contour). In this picture, the dynamical transition of ignition is to cross the expanding separatrix and begin following a closed contour (middle). As the separatrix expands further, the system remains on an $H_{\mathrm{eff}}$ contour whose enclosed area stays constant (adiabatic theorem).\\
\textbf{c)} A one-parameter family of trajectories in phase space (blue surface), initially all on unbound orbits, evolves in time (vertical axis). The wireframe is the surface through which any horizontal slice is the closed lobe of the separatrix at a given time; the three slices shown correspond to the planes in b). The green spiral curve is the example trajectory shown in b) and a). As the wireframe separatrix expands, incompressible flow requires that a finite fraction of the initial trajectories is inhaled.
\\
\textbf{All:} The dynamics plotted is for $V(q,U) = -\sqrt{1- U^{2}}\cos(q)$ with no $\vec{\alpha}$, $(\delta,\epsilon,g) = (0.2,0.1,0.2)$, and $J=1.481$. The three successive times for plots and planes are (bottom to top) $t_1=4.25$, $t_2=10.625$, $t_3=17$. Initial conditions for the green example trajectory are $q(0)=0.575\pi$, $p(0)= 0.597$. The sheaf of initial trajectories in c) all have equal $H_{\mathrm{eff}}$ at $t=0$.}
\end{figure*}

Along with conservation of total energy, the universal equations (\ref{eq:Canon}) imply a second identity that is equally fundamental: Liouville's theorem, which states that the motion of all points in phase space is an incompressible flow. Therefore, let $S(\epsilon t)$ be the phase space volume of bound orbits around a local minimum of the time-dependent $V_\mathrm{eff}$. $S$ is straightforwardly computable for any specific instantaneous $V$. If $S$ is increasing with $\epsilon t$, then some previously unbound orbits \textit{must} be drawn into this volume, becoming bound. Precisely which orbits will thus begin performing downconversion depends on detailed evolution of high-frequency degrees of freedom, but the fraction of initial states that lead to downconversion can be computed in terms of derivatives $dS/dt$. Conversely, if $S$ is \textit{shrinking}, then \textit{no} unbound orbits can become bound, for incompressible flow cannot squeeze more orbits into less space.

With a daemon whose particular $V$ provides $\dot{S} > 0$, one may simply launch the weight upward at any speed $\dot{z} > 1/\delta$, and let gravity slow it. When the critical speed $\dot{z} = 1/\delta$ is reached, the daemon engine will sometimes (depending on detailed initial conditions) ignite: $\dot{z}$ will cease its steady falling and commence oscillating around $1/\delta$. See Fig.~(2). Numerically solving (\ref{eq:Canon}) in such cases, choosing initial conditions at random, is uncannily reminiscent of striking a match or pulling a lawnmower cord. Which attempt will succeed can depend sensitively on initial conditions, but the chance of success is sufficient to simply keep trying. It can even reach 100\%. The dynamical transition to steady downconversion is thus \textit{spontaneous} in both colloquial senses of the word. Without knowledge of high-frequency degrees of freedom, it is unpredictable; without control over high-frequency degrees of freedom, it still occurs.

If  $S$ is not increasing, in contrast, then even though $V_\mathrm{eff}$ allows bound orbits, when one launches the weight upward at initial $\dot{z}>1/\delta$, the daemon \emph{never} ignites. \textit{The necessary and sufficient condition for autonomous daemon ignition is}
\begin{equation}\label{Micro2ndLaw}
\frac{dS}{dt} > 0\;. \end{equation}
In the Supplementary Material we provide concrete cases of $V$, whose equations of motion can be solved on any laptop computer, to show that this condition is non-trivial (the simplest forms of $V$ do not satisfy it) but achievable with simple forms of $V$, including the system illustrated in Fig.~2, whose bilinear coupling between two high-frequency oscillators makes it resemble a Carnot engine running between hot and cold reservoirs. Condition (\ref{Micro2ndLaw}) represents an unavoidable but manageable design constraint on daemon engines --- and offers a robust and general opportunity, based on Liouville's theorem, to exploit high-frequency motion without detailed control over it. 

We have thus shown that classical mechanics permits steady downconversion by autonomous microscopic engines, and even allows these daemon engines to ignite autonomously, without fine control over high-frequency variables. The mechanism of daemon operation is basic and general: a velocity-specific resonance can be self-sustaining around a local minimum of the nonlinear coupling potential. Daemon stalling and spontaneous ignition are also universally constrained by adiabatic invariance and Liouville's theorem. While the constraints on daemons are reminiscent of thermodynamic limits, however, they have been obtained within closed-system classical mechanics without chaos or disorder. The mechanical limits and opportunities for daemons thus do not seem identical to those of statistical mechanics or macroscopic thermodynamics; and they apply in a regime to which those theories are not thought to apply. 

The regime in which daemons can operate requires further study. Although by definition \emph{able} to operate as closed systems, daemons operate on robust principles of adiabatic theory, and may therefore be expected to tolerate dissipation and noise. If so, they may one day be realized in practical microscopic devices that use nonlinear resonances to sustain steady downconversion and run for long times on densely stored internal energy --- or even identified within the natural molecular machinery of living organisms\cite{biomotor2}.

\section{Acknowledgements}
LG acknowledges funding from the German Excellence Initiative (DFG/GSC 266). 

\clearpage

\newpage

\onecolumngrid
\setcounter{equation}{0}
\setcounter{figure}{0}
 \renewcommand{\theequation}{S-\arabic{equation}}
 \renewcommand{\thefigure}{S-\arabic{figure}}

\textbf{\Large Supplementary Material}\\

Our Supplementary Material consists of four Sections and an Appendix. The derivation of our basic daemon engine Hamiltonian (Eqn.~(2) in the main text) is Section~1; Section~2 is a brief review of Liouville's Theorem; and an explanation of the non-trivial consequences for daemons of $U$ being bounded from below is provided in Section~3. Section~4 supplies a comparison of two specific daemon models, which shows how the universal mechanical constraints on daemons take specific form for specific systems, and demonstrates that the mechanical limits on microscopic engines are non-trivial but manageable: both daemon engines can run, but only one can start autonomously. The Appendix presents formal multiple scale analysis to justify some statements about resonance that are crucial to Section~1.

\section{Daemon Hamiltonian}

In this discussion we revert from our main text's dimensionless variables to ordinary, dimensionful ones. We begin by explaining why a daemon Hamiltonian should in general have two small parameters $\delta$ and $\epsilon$. We then explain why the special functional form $V(z,\tau,U,\vec{\alpha})\to V(z-\tau,U,\vec{\alpha})$ may be assumed, with broad generality, for a daemon which can lift the weight at a given steady speed.

\subsection{Small parameters}
If all relevant degrees of freedom are represented (including those usually regarded as reservoir or environment), any engine can be described by a Hamiltonian of the form
\begin{equation}\label{eq:Hgen}
H = U + W + V\;,
\end{equation}
where $W$ is the energy of the system on which the engine may do work, $U$ is the energy of fuel and any other high-frequency degrees of freedom, and $V$ describes the interaction that can transfer energy from $U$ to $W$. 

\begin{figure}[htbp]\label{fig:sketch}
\centering
\includegraphics[width=.4\textwidth]{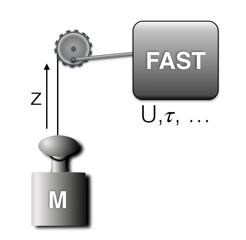}
\caption{
A general daemon engine as a closed system within which energy from high-frequency degrees of freedom can perform steady work.}
\end{figure}

Applying Carnot's premise that any work is equivalent to lifting a weight, we take $W$ to describe a mass $M$, with height and upward momentum $(z,p)$ as its canonical pair of dynamical variables, carrying ordinary kinetic energy and subject to a constant downward force $-Mg$:
\begin{equation}\label{eq:HW}W = \frac{p^2}{2M} + Mg z \;.\end{equation}
The restriction to an exactly constant external force is an unimportant convenience, whose straightforward generalization we will discuss below. In order to ensure that the weight is truly an ordinary weight, with momentum $p/M = dz/dt = \partial H/\partial p$, and not some more exotic kind of degree of freedom, we stipulate that neither $V$ nor $U$ can depend on $p$. The behavior of a daemon system will depend dramatically on the weight's velocity, but not because of any direct coupling to the weight's momentum.

We wish to consider a daemon engine which will perform steady downconversion, like macroscopic engines, but do so with only a few degrees of freedom, and without dynamical chaos. We exclude chaos simply to show that engine-like behavior is possible within very simple dynamical systems; extending our theory to encompass chaotic daemons will be an important goal for future work. Since the fuel subsystem is thus by assumption integrable, the Hamilton-Jacobi construction lets us describe it without loss of generality using its internal energy $U$ itself as one canonical momentum (filling the role, say, of $p_1$), whose conjugate co-ordinate we denote as $\tau$ (which thus plays the role of $q_1$). We refer to all other fuel system dynamical variables $(q_{n>1},p_{n>1})$ as $\vec{\alpha}$ for compactness. Within this special case of integrable high-frequency subsystems, the abstract Hamilton-Jacobi representation is otherwise very general; the sector which is represented by $(\tau,U,\vec{\alpha})$ might include integrable microscopic analogs (insofar as such are possible) for hot and cold reservoirs, exhaust and waste products, etc. They could even represent working parts in the daemon engine itself, if any of these are for some reason independently movable, rather than being constrained to follow the evolution of $z$ or $\tau$. We therefore consider $V = V(z,\tau,U,\vec{\alpha})$. This still leaves $H$ as a rather general Hamiltonian; we now consider what special forms of $V$ may let it represent a daemon.

Firstly the daemon should be small; hence $V$ should depend on the weight height in relation to a short characteristic length scale $\lambda$. In other words $V$ should depend on $z$ only in the dimensionless form $\tilde{z} = z/\lambda$, and without any large or small factors involved in this $\tilde{z}$-dependence. (In subsection 1.4, below, we will address the possibility of there being more than one $\lambda$ scale involved in the dependence of $V$ on $z$.)

Secondly the daemon should genuinely be a microscopic engine, and not just mathematically disguised clockwork. That is, the daemon must perform downconversion, transferring energy from high-frequency modes into much slower motion. Purely mathematical rescaling can make any variable formally slow or fast, but if we insist that the physically accessible variables on which a daemon's interactions can directly depend should be high-frequency ones, then what this means is that the abstract Hamilton-Jacobi co-ordinate $\tau$ must be some linear combination of fast accessible variables, and hence $V$ can only depend on $\tilde{\tau}=\Omega\tau$ for some high frequency $\Omega$. It will turn out that we can identify the speed with which the daemon lifts the weight as $v_c = \lambda\Omega$; we will simplify our expressions by assuming this scale relationship now and confirming it below. (In subsection 1.4 we will also discuss the possibility of there being more than one scale $\Omega$.)

Thirdly the daemon should benefit from downconversion by being able to do a lot of work in proportion to its size. In particular it should certainly be able to do more work by steady downconversion than the energy needed to accelerate it to running speed, and so if we write $U = U_0 \tilde{U}(t)$ with initial $\tilde{U}(0)$ of order 1, then we should have $Mv_c^2 = U_0\eta$ for $\eta\lesssim 1$. 

Finally, a bomb that rapidly accelerated the weight with its blast might be achieving downconversion, but a daemon's downconversion should be \textit{steady}. The characteristic energy scale of the daemon interaction $V$ should therefore not be the rising weight's entire kinetic energy $Mv_c^2$, but some much lower energy associated with sustaining steady motion against the external force $-Mg$. Hence we assume $V = V_0\tilde{V}$ for bounded dimensionless $|\tilde{V}|\lesssim 1$ and $V_0 = \delta^2 Mv_c^2$ for $\delta\ll 1$. The steadiness of daemon dynamics should also mean that the interaction changes only gradually as $U$ decreases, and so $V$ should depend on $U$ as a function of $U/U_0$, and should do so gradually, rather than with a steepness that could be characterized by a further small parameter.

If we now convert entirely to dimensionless variables $\tilde{t} = t\Omega\delta$, $\tilde{H} = H/(U_0\delta)$, $\tilde{p}= p\,v_c/U_0$, and $\tilde{g} =  Mg\lambda / V_0$, and define $\epsilon = \eta\delta\ll 1$, then the canonical form of the Hamiltonian equations of motion (1) is preserved for the dimensionless problem, with evolution in $\tilde{t}$ instead of $t$ itself, by
\begin{eqnarray}\label{tildeH}
\tilde{H} = \frac{\tilde{p}^2}{2\epsilon} + \frac{\tilde{U}}{\delta} + \epsilon\tilde{g}\tilde{z} + \epsilon\tilde{V}(\tilde{z},\tilde{\tau},\tilde{U},\vec{\alpha})\;.
\end{eqnarray}
The appearance of two necessarily small parameters $\epsilon$ and $\delta$ in the basic daemon Hamiltonian reflects the challenging nature of the daemon's defining task, of converting a large amount of energy from high-frequency motion into work on a weight, yet doing so steadily rather than violently. Daemons are a special class of system.

To confirm that specifying these two parameters as small really captures what it should mean for a dynamical system to be an `engine' in a meaningful sense, we can consider a typical automobile engine.  It uses perhaps six liters of fuel per 100 kilometers, while driving at 100 km/h, hence consuming on the order of 0.1L of fuel per minute. Gasoline yields around 40 MJ per liter, so the car engine takes in about 4 MJ per minute; at 4000 RPM, this implies that typical car engine consumes about a kiloJoule of chemical energy per cycle. Since this energy is mostly injected as potential energy at a moment when the piston is not moving, it represents the overall energy scale of the dynamical coupling between fuel and vehicular motion. If cars were somehow driven by daemons, therefore, we would estimate them to have $V_{0}\sim$ 1 kJ. Comparing this scale to $Mv_{c}^{2}\sim 1MJ$ for a 1500 kg car traveling at 100 km/h, we find $\delta = \sqrt{V_{0}/(Mv_{c}^{2})}\sim 0.03$. Comparing $Mv_{c}^{2}\sim$ 1 MJ to the roughly 2000 MJ of energy in the car's full fuel tank then implies $\eta\sim 5\times 10^{-4}$ and hence $\epsilon=\eta\delta\sim 10^{-5}$. 

Lowering either $\delta$ or $\epsilon$ well below the above figures would produce a car engine whose efficiency or range would let it dominate the market; while to raise either of them to near unity would produce a machine which no-one who wanted an engine would ever buy. Practical applications of microscopic engines are unlikely to be exactly like those of automobile engines, but it is indeed the smallness of $\epsilon$ and $\delta$ which expresses quantitatively the practical advantages of downconversion over single-time-scale clockwork for supplying steady power compactly. In strictly pragmatic terms, therefore, the regime of small $\delta$ and $\epsilon$ is the regime of any useful microscopic engine.

A third potentially significant parameter for a daemon is $\tilde{g}$. If it is large, the daemon will not work, because the maximum force which the daemon can exert on the weight is limited by the scale of $\epsilon\tilde{V}$, and if the external force $\sim \epsilon\tilde{g}$ exceeds this then the daemon can never lift the weight. Daemon engines do not really differ in this respect from macroscopic engines, which also have torque or force limits as well as power and efficiency limits.

In summary, the basic form of Hamiltonian that we assume in the main text represents the general daemon case of work as lifting a weight, using integrable high-frequency fuel, with small parameters ensuring that the engine can deliver more work than it takes to get it running, and do so steadily rather than violently. 

\subsection{Effective Hamiltonians}

We now show why we can restrict the heretofore general dependence of $V$ on $z$ and $\tau$ to dependence on the difference $z-\tau$,
\begin{equation}
\tilde{V}(\tilde{z},\tilde{\tau},\tilde{U},\vec{\alpha})\longrightarrow\bar{V}(\tilde{z}-\tilde{\tau},\tilde{U},\vec{\alpha})\;.
\end{equation}
This is a technical issue, but crucial for our entire paper. We will discuss it here in terms of the familiar basic concept of resonance, and then in the Appendix to this Supplementary Material we will use multiple scale analysis to explain what precisely resonance means, in technical terms, in a nonlinear dynamical system.

Let us first note that, because of the small coupling parameter $\epsilon$ and the stipulated mismatch of dynamical time scales ($\tilde{t}$ versus $\tilde{t}/\delta$), the default expectation is that the effects of $\tilde V$ on system dynamics would simply be negligible. The conclusion would then be that $U$ and $W$ would essentially be separately conserved, with $\tilde V$ giving rise to no more than small variations in either. Such \emph{adiabatic decoupling} is indeed not only an excellent approximation in typical cases, but an essential and basic tool throughout all of theoretical physics, represented in effective Lagrangians, adiabatic elimination, renormalization, and all methods which assume that high-frequency degrees of freedom may modify slow time evolution, but will not steadily inject energy into the slow sector (or remove energy from it). Adiabatic decoupling is even familiar from school physics, as the principle that weakly coupled systems do not (significantly) affect each other unless they are in resonance. 

As we will see, however, the particular form of $\bar{V}$ represents an interaction that does \emph{not} necessarily adiabatically decouple; and so while other forms of $V$ may be neglected, $\bar{V}$ must be retained. The assumption in our main text is that this step of neglecting non-resonant terms in $V$ has already been made, leaving only a term of the special $\bar{V}$ form. In assuming the $\bar{V}$ form that we use for a daemon, we are therefore not so much restricting ourselves from the case of general $V$, as generalizing beyond the default assumption that $V\to0$ effectively. 

As a concrete example, consider a simple case in which the high-frequency subsystem is a single harmonic oscillator, with no additional $\vec{\alpha}$ degrees of freedom:
\begin{eqnarray}\label{HOengine}
U &=& \frac{1}{2}P^2 + \frac{\Omega^2}{2}Q^2\nonumber\\
V &=& kQ\cos(z/\lambda)
\end{eqnarray}
for conjugate canonical variables $Q$ and $P$ and some coupling constant $k$. In Hamilton-Jacobi variables the harmonic oscillator co-ordinates are expressed as
\begin{eqnarray}\label{HJOsc}
Q &=& \frac{\sqrt{2U}}{\Omega}\cos\Omega\tau\nonumber\\
P &=& -\sqrt{2U}\sin\Omega\tau\;,
\end{eqnarray}
so that we can re-write the above simple $V$ in our $U,\tau$ terms as
\begin{eqnarray}\label{HJOsc2}
V &=& k\frac{\sqrt{2U}}{\Omega}\cos(\Omega\tau)\cos(\lambda^{-1}z)\nonumber\\
 &\equiv&k\frac{\sqrt{U}}{\sqrt{2}\Omega}[\cos(\lambda^{-1}z-\Omega\tau) + \cos(\lambda^{-1}z+\Omega\tau)]\nonumber\\
&\to&\epsilon\sqrt{\tilde{U}}[\cos(\tilde{z}-\tilde{\tau})+\cos(\tilde{z}+\tilde{\tau})]
\end{eqnarray}
when in the last line we adopt our dimensionless variables as described above. 

The resonance condition that is important here concerns the matching of the temporal frequencies of the two factors $\cos(\tilde{\tau})$ and $\cos(\tilde{z})$. Since $d\tilde{\tau}/d{\tilde t}=1/\delta+\mathcal{O}(\epsilon)$, these frequencies will be close whenever $d\tilde{z}/d\tilde{t} \doteq \pm 1/\delta$. Whereas ordinary linear resonance concerns the matching of fixed motional frequencies, in this case the resonance occurs when the weight itself has no frequency, since its motion is not periodic; here the resonance instead occurs when the weight moves steadily at a certain special speed (either up or down). When the weight moves at either of these speeds, one of the $\cos(\tilde{z}\pm\tilde{\tau})$ terms in (\ref{HJOsc2}) will be constant (or very nearly), while the other will oscillate rapidly. If the weight's speed is not resonant, then both terms will oscillate rapidly. The principle that weak couplings can only have significant effects when they are resonant is simply the observation that a rapidly oscillating term in $V$ cannot have steadily accumulating effects. For a technical demonstration that this is so, using multiple scale analysis, see the Appendix at the end of this Supplementary Material. 

In the example of this particular $U$ and $V$, we can directly confirm the validity of the resonance principle. The exact evolution under the full $H$ can easily be solved numerically, and compared with the evolution obtained using the $H_{\mathrm{eff}}$ obtained by discarding the $\cos(\tilde{z}+\tilde{\tau})$ term from $V$. When $\delta$ is small, the two evolutions are indeed extremely close --- except where $d\tilde{z}/d\tilde{t}$ is close to \textit{minus} $1/\delta$. In that range of weight speeds, it is the $\cos(\tilde{z}+\tilde{\tau})$ term which can be significant, because it is nearly constant, rather than the $\cos(\tilde{z}-\tilde{\tau})$ term, which oscillates quickly. The significant effect which the $\cos(\tilde{z}+\tilde{\tau})$ term can have, when the weight falls at the resonant speed, is to keep the weight \textit{falling} at close to the constant speed $-1/\delta$, despite gravity; downconversion operates in reverse (upconversion) and instead of a daemon engine we have a daemon brake. 

Daemon brakes can be analyzed in very similar ways to daemon engines, but we do not discuss them here. If the weight is falling at a speed near $-1/\delta$, then it will simply keep on falling, whether at nearly constant speed (because of daemon braking) or accelerating downward as usual under gravity; but for all of phase space apart from the narrow range of speeds around $-1/\delta$, including all cases where the daemon engine does run, it is sufficient to retain only the $\cos(\tilde{z}-\tilde{\tau})$ term in the dimensionless $V$. In this sense, only the $\cos(\tilde{z}-\tilde{\tau})$ is necessary to describe a daemon engine that operates by this example interaction (\ref{HJOsc2}).

(One somewhat subtle point is important to appreciate here: to retain $\bar{V}$ in $\tilde{H}$ is \emph{not} to assume that the weight actually is rising at the special speed. Moreover, even if the weight does happen to be rising at the special speed, retaining $\bar{V}$ in $\tilde{H}$ does not assume or ensure that $\bar{V}$ actually will have significant effects. The actual effect of the retained $\bar{V}$ term remains a question of time evolution in Hamiltonian mechanics. Daemon engines can stall, or fail to ignite, as dynamical transitions occur or do not occur. If the actual effect of the retained $\bar{V}$ term should turn out to be slight, then it would be inconsistent to analyze just what tiny effect it did have, without also including the comparably tiny effects of neglected terms. If one merely concludes that no significant effects have occurred, as we do in such cases, then this is still an accurate conclusion as far as it goes, and there is no need to take those neglected terms into account in order to confirm it.)

So, we discard non-resonant terms because they will never have significant effects in any region of phase space in which we are interested, and we retain $\bar{V}(\tilde{z}-\tilde{\tau},\tilde{U},\vec{\alpha})$ terms because they possibly might have significant effects near one special weight speed. The technical point to make in this connection is that the $J$ of our main text's Eqn.~(3) need not really be an exact constant of the motion, as it is if $V$ really has exactly the $\bar{V}$ form. If the actual $V$ also contains non-resonant terms, which are completely ignored in the main text, then $J$ will merely be an \textit{adiabatic invariant} rather than an exact constant: it may actually have small, high-frequency oscillations, but these will remain small and rapid, and there will be no steady change in $J$ over intermediate or even long times. The role of $J$ in constructing the reduced $H_\mathrm{eff}$ will be the same, but the resulting $H_\mathrm{eff}$ will be an \textit{adiabatic effective Hamiltonian} instead of an exact Hamiltonian for the low-frequency motion. Such an adiabatic effective Hamiltonian will approximate the exact evolution of the system, but only by leaving out oscillations and fluctuations that are both small in amplitude and high in frequency.

Just as we claimed, therefore, replacing $\tilde{V}(\tilde{z},\tilde{\tau},\tilde{U},\vec{\alpha})\to\bar{V}(\tilde{z}-\tilde{\tau},\tilde{U},\vec{\alpha})$ is really a matter of being more general than the default assumption of adiabatic decoupling of weakly coupled subsystems with greatly differing time scales, which would be the adiabatic replacement $V\to 0$. With $V=0$ we would have $U$ and $p+\epsilon g t$ both conserved separately; if $V\to0$ were a valid adiabatic approximation, then we would have $U$ and $p+\epsilon g t$ as two independent adiabatic invariants. It is a more general approach, and not a more special assumption, to take only the sum $J$ of these two to be adiabatic invariant \textit{a priori}, and allow dynamical evolution under $H_\mathrm{eff}$ to determine whether or not the two parts of $J$ really do remain separately constant up to small oscillations.

It is thus a curious but essential feature of daemons as dynamical systems that they possess such an apparent oxymoron: an explicitly time-dependent adiabatic invariant, $J=U+p+\epsilon g t$. The apparent oxymoron is not really contradictory. The explicit appearance of $t$ in $J$ just means that the sum $U+p$ must change steadily, in order to keep $J$ constant. This simply reflects conservation of total energy. Given gravity, if the weight does not accelerate downward, then fuel must be expended. Moreover, the reason $J$ is adiabatically invariant is precisely because $\delta$ is small; that is, because high-frequency motion is being tapped for slow work. 

The most basic effect of $\delta$'s smallness is to make the lifting speed large in microscopic time and length units ($\lambda\Omega\to1/\delta$), so that the daemon produces a large amount of power for its size, as an engine should. The inextricable secondary effect of small $\delta$, however, is to make the explicitly time-dependent $J$ to be adiabatically invariant. This then leads, upon standard adiabatic elimination of the highest frequency dynamics, to the explicitly time-dependent effective Hamiltonian $H_\mathrm{eff}$, even though the entire system is closed and the full $H$ is time-independent. Since the highest frequencies have been eliminated, $\delta$ then plays no further direct role; but its vital work has been done. $H_\mathrm{eff}$ is time-dependent even though --- indeed, precisely because --- the whole system is closed, the total $H$ is time-independent, and the total energy is exactly conserved. 

In particular, we emphasize again: the time-dependence of $H_\mathrm{eff}$ is not put in by hand phenomenologically, to represent the assumed fact that the daemon performs work. On the contrary, $H_\mathrm{eff}$ is a valid adiabatic effective Hamiltonian to describe the intermediate and long-term dynamics of the system, whether or not the daemon engine runs; and it is therefore able to describe the daemon's starting and stopping, as dynamical transitions.

Readers who wish to see exactly what resonance and frequency really mean, in highly nonlinear dynamics, may examine the Appendix at the end of this Supplementary Material, which applies multiple scale analysis to daemon dynamics, to show that terms in $V$ which are not slow when the weight moves at the lifting speed can have only small effects on time evolution, even over long time scales.

\subsection{Generalizations}

We now discuss some possible generalizations of the scenario we assume in the main text. Some of these can in fact be described by simple adaptations of our theory as presented. Others may require further extensions of the theory.

It is possible to consider daemons doing work against non-constant forces, by generalizing our model to allow $\tilde{g}$ to depend slowly on $\tilde{z}$. In numerical studies of such models we have confirmed that they can be described adiabatically, in the sense of just applying the constant-$g$ theory explained in this paper, using whatever external force applies at the weight's instantaneous position as the instantaneous $-Mg$. This includes the possibility of an external force that gradually becomes stronger, as in a car driving up a steepening hill. If the external force strengthens to the point where it exceeds the daemon's force limit, then the daemon engine will stall. This limitation on daemons is not a different one from the phase-space volume limit we explain in our paper, however; it is merely another case of the same, because a strengthening external force tends to shrink the phase space volume within which instantaneous periodic orbits exist. In this sense, any slowly varying external force is essentially equivalent to the exactly constant one we consider explicitly. To treat more abruptly varying external forces will require fresh analysis, although at some point a rapidly varying external potential may make it dubious whether $W$ really counts as slow work in the sense we have assumed as part of the definition of a daemon.

We can also ask what happens if the daemon $V$ depends on $z$ and $\tau$ through more than one length scale $\lambda$ and frequency $\Omega$. In a minimal sense this may occur generically, in that that the integrable fuel sector may be described by action-angle variables, such that $U$ is a (possibly nonlinear) function of the action variables and $\tau$ is some (possibly action-dependent) linear combination of the angles. A daemon which coupled directly to the angle variables could in general be expressed as coupling to $\tau$, as in our main text; but $\Omega$ would then in general depend on the action variables, so that $\Omega\to\Omega(U,\vec{\alpha})$, and hence to a slowly time-dependent $\Omega$ in general. As long as this time-dependence is slow, however, the theory of our main text may be applied adiabatically, with the weight-lifting speed slowly changing. 

(Ensuring sufficient slowness of $\Omega(U,\vec{\alpha})$ requires that the nonlinear dependence of $U$ on action variables be sufficiently weak, in comparison to the coupling perturbation $\epsilon V$. The fact that this condition is opposite in direction to the isoenergetic nondegeneracy condition assumed by the Kolmogorov-Arnol'd-Moser theorem suggests that daemon dynamics lies outside KAM theory. In any case, the dynamical transitions we describe as daemon ignition must not be forbidden by any theorem, because it is easy to verify by numerical solution using well-tested commercial software tools that the evolution of the rather simple systems that we describe truly is as we describe.) 

One could also simply have a coupling with multiple distinct $\Omega$ at any time. This would provide multiple lifting speeds $\lambda\Omega$. If $V$ offers lifting speeds differing only by order $\epsilon$, then this is equivalent to having a $V$ with one lifting speed, and some additional slow time dependence, such as our theory already considers. If the lifting speeds differ by significantly more than order unity in our dimensionless units --- which may still mean differing by a small fraction of $1/\delta$ --- then these resonances will be far enough apart in phase space that they can simply be considered separately. Ignition, running, and stalling near each speed will be described by our theory; if the weight later approaches a different lifting speed, the theory for a daemon with \textit{that} lifting speed will apply --- after re-scaling dimensionless variables and redefining small parameters to set the new lifting speed equal to the new $1/\delta$.

A system with resonant speeds differing by an amount that is neither large nor small (in our dimensionless units) may require fresh analysis, however. Suppose we had
\begin{eqnarray}\label{twores}
V = V(\lambda^{-1}z-\Omega\tau,\lambda^{-1}z-(\Omega+\omega)\tau,U) \to \tilde{V}(\tilde{z}-\tilde{\tau},\tilde{z}-\tilde{\tau} - \delta\,\tilde{\tau},\tilde{U})\;.
\end{eqnarray}
with $\omega$ of order $\delta\times\Omega$. When $d\tilde{z}/d\tilde{t}\doteq 1/\delta$, so that the first argument of the dimensionless $\tilde{V}$ is slowly changing (resonant), the second argument will be $\sim \tilde{t}$, implying an effective potential that is time-dependent on a time scale comparable to the period of bound orbits featuring downconversion, if these exist. In many cases such time-dependent perturbations may in fact simply prevent downconversion by knocking the system out of bound motion; one might say in such cases that overlapping nonlinear resonances tended to disrupt each other. Driven systems \textit{can} exhibit bound motion, however, and so we cannot rule out a class of more complex daemons to which our simple theory does not apply. Since $H$ will still be conserved in these cases, steady lifting of the weight in the external potential will still require steadily dropping $U$, so that $V$ slowly changes; and Liouville's theorem will apply in these cases as well. We therefore expect that a generalized form of our constraints will also limit this more complex form of microscopic engine.

Exploring such generalizations may be a good first step towards filling in the gap between daemon dynamics and thermodynamics, because a microscopic engine with multiple overlapping resonances is already a step closer to macroscopic engines. If for example a macroscopic piston engine were to use a strictly non-interacting gas as a working fluid, each collision between a molecule and the heavy piston would be nearly elastic, and the molecule might well shift between periodic bouncing orbits in the cylinder with only slightly different periods. The ultimate extreme downconversion from molecular bond frequencies to crankshaft turn rates would proceed through a long cascade of very many small frequency steps. Experience amply shows that such cascades can work, and in this sense one can say that the limit of a daemon engine with very many overlapping resonances \emph{is} a macroscopic engine.

Such complexity has hitherto normally been considered necessary for a thermodynamical engine. Here we are pointing out that much simpler engines are in principle possible, and yet even these must still obey limits that are surprisingly analogous to thermodynamics. The theory of daemon engines that run on \emph{isolated} nonlinear resonances is the topic of this paper. It establishes the possibility of daemons as dynamical systems, but shows that they obey inherent non-trivial limits beyond energy conservation alone. The question of systems that interpolate between simple daemons and macroscopic engines is a natural one, which might have profound implications for the microscopic roots of thermodyamics itself; but it is a question beyond the scope of this paper.

One class of daemons with multiple time scales is already included in our theory, albeit implicitly: one might include additional degrees of freedom with dimensionless time scales on the order of unity, so that after eliminating the fastest modes and treating slow variables adiabatically, the dynamics of the daemon engine cycle would still take place in a phase space of more than two dimensions. If these extra degrees of freedom are parts of the daemon engine, then its small size should imply that their phase space is compact, so that the total phase space volume of bound orbits remains finite, and the theory of the main text can be straightforwardly applied --- though exactly what its application has to say about such daemons is also a subject for further research. 

For a daemon engine's coupling mechanism to possess its own degrees of freedom, apart from the weight and the fuel, is a non-trivial complication. Simpler models which lack such extra engine degrees of freedom do not represent systems with no engine components; they merely describe systems in which no engine components are free to move independently of the weight and fuel. It is not clear that allowing such independent motion would be a good idea for daemon engines. After all, macroscopic engine designs do not normally specify loose parts that are free to rattle at the same frequency as the crankshaft. It might possibly be useful, however, to include extra degrees of freedom in a daemon so that they can serve as a sink for action, and allow the area enclosed by the system's orbit in the $q,p$ plane to decrease as the daemon engine runs. This could delay stalling or assist ignition.

One could regard this kind of daemon complication as an attempt to incorporate a Hamiltonian Maxwell's Demon into the daemon engine, in order to break the microscopic analog of the Second Law (equation (6) of our main text). In fact the downconversion volume law (6) would not break, because it is expressed in terms of a (possibly multi-dimensional) phase space volume, which in this kind of case would be a volume in the total phase space of all dynamics on the unity time scale. If additional degrees of freedom absorbed action from $q$ and $p$, this would only mean that the total volume of bound orbits $S$ was larger. It is conceivable, however, that Hamiltonian `demons' might prove to be components comparable to turbochargers, able to improve daemon efficiency by improving the behavior of $S$ as a function of $U$.

\section{Energy conservation and incompressible flow}
Two universal identities follow from the fundamental Hamiltonian equations (1):
\begin{align}\label{IDs}
&\text{I.}  &\frac{d H}{dt}  \equiv \frac{\partial H}{\partial t} + \sum_n \left[\frac{\partial H}{\partial q_n}\frac{d q_n}{dt} + \frac{\partial H}{\partial p_n}\frac{d p_n}{dt} \right] &= \frac{\partial H}{\partial t} + 0\nonumber\\
&\text{II.} & \sum_n \left[\frac{\partial}{\partial q_n}\frac{d q_n}{dt} + \frac{\partial}{\partial p_n}\frac{d p_n}{dt}\right] &= 0\;.
\end{align}
Identity I means that, as long as $H$ does not depend on time $t$ directly, but only through its dependence on the time-dependent $q_n(t)$ and $p_n(t)$, then the value of $H$ will not actually depend on $t$ at all, because the $q_n(t)$ and $p_n(t)$ will only change in ways that maintain a constant value of the combined function $H$. 

The value of $H$ is called the total energy of the system, whatever $H$ or the system may be. Fundamental Hamiltonians do not depend on $t$ directly, and hence total energy is conserved. Further analysis or approximation may introduce effective Hamiltonians which only partially or indirectly describe the entire system, and these may depend directly on $t$. Daemons are a case in point. Such a time-dependent effective Hamiltonian is in general not conserved, although the full energy always is. In the daemon case, for example, $H$ is always exactly conserved but $H_\mathrm{eff}$ is not, because energy can be exchanged between the low-frequency motions which are represented in $H_\mathrm{eff}$ and the high-frequency motions which are not.

Identity II is Liouville's theorem, and it applies to all Hamiltonian systems, even if $H$ depends directly on $t$. If one considers $\vec{r} = (q_1, q_2, ..., p_1,p_2,...)$ as a point in (generally) multi-dimensional phase space, then each possible $\vec{r}$ is a set of initial conditions sufficient to determine future evolution under classical mechanics. In particular $d\vec{r}/dt$ is determined, for every such $\vec{r}$, by the Hamiltonian equations of motion (1): where the system will go next in phase space is uniquely determined by where the system is now. This is mechanical causality; future and past are mapped to each other one-to-one. What Liouville's theorem further states, however, is that the vector field $d\vec{r}/dt$ is divergenceless.

Liouville's theorem thereby means that an especially strong form of causality is built into mechanics at the same fundamental and universal level as conservation of energy. The mapping between each moment and the immediate future is not just one-to-one: it is an incompressible flow in phase space. Time evolution cannot tend, even in the slightest, to converge or diverge, concentrate or diffuse. If one considers any phase space volume of different initial conditions $\vec{r}$, and follows all those alternative histories of the system through time, then their collective phase space volume will be conserved exactly.

Even though Liouville's theorem is just as fundamental as energy conservation, it is less often considered. This may be because Liouville's theorem is a strangely counterfactual condition, inasmuch as it says nothing directly about how any single system will evolve in any single case. In classical mechanics, the state of any system at any instant is a point in phase space. Volume conservation does not constrain how that point moves over time, because the volume of a point is always zero, no matter where the point goes. Rather, Liouville's theorem constrains the ways in which hypothetical alternative points, beginning near each other, could move in relation to each other. It is a law about paths not taken; a rule about what could happen rather than about what does. Liouville's theorem can nonetheless provide useful information. One can sometimes deduce non-trivial facts about what will happen from information about what could happen.

The universal mechanical basis for the First Law of Thermodynamics has long been clear, in total energy conservation. What thermodynamics adds to mechanics is mainly the Second Law, about entropy increase in spontaneous processes. It seems fair to expect that if something as profoundly and universally important as the Second Law is to be an emergent property within pure mechanics, then it should ultimately be due to some significant basic principle, and not to a mere coincidence of details. At the level of fundamental mechanics, Liouville's Theorem is the only obvious candidate to be the basis for the Second Law. All of statistical mechanics already depends on Liouville's Theorem, inasmuch as conservation of probability in ensemble theory depends on it. Here we have shown that Liouville's theorem implies mechanical conditions on spontaneous daemon ignition which are remarkably similar to the thermodynamic condition of entropy increase. This suggests an interesting new perspective on the microscopic origin of thermodynamics: perhaps thermodynamics does not emerge from mechanics in the macroscopic limit, but instead simply persists into the macroscopic limit, as a set of constraints that apply to all processes of steady downconversion, and as such are also fully valid microscopically. 

\section{Consequences of $U$ being bounded from below}

Infinite energy sources are unfortunately unavailable; every possible $U$ must be bounded from below. What this means is that there must exist some value of $U$ from which no smaller value of $U$ can evolve, and which cannot have evolved from any smaller value of $U$. At this value, therefore, $dU/dt = 0$ must hold, for all values of other dynamical variables. This property of the full and exact dynamics must be shared, at least approximately, by the approximate dynamics governed by $\bar{H}$; the approximation might perhaps slightly shift the minimum value of $U$, but it must preserve the qualitative feature that such a minimum value exists.

From the canonical equations of motion this means that at the minimum value of $U$ according to $\bar{H}$ we must have $\partial \bar{V}/\partial_\tau = -\partial\bar{V}/\partial q = 0$. At this minimum value of $U$, therefore, if not already at some higher value, $V_\mathrm{eff}$ must cease to have any local minima at all, and the phase space volume of bound orbits must be zero. At least within some neighborhood of this minimum $U$, therefore, the phase space volume of bound orbits must be shrinking with decreasing $U = J - p -\epsilon g t$, and hence shrinking with increasing $t$. 

\section{Growing separatrix volume as a manageable constraint}

The condition of growing separatrix volume, in order for a daemon to ignite spontaneously, is non-trivial. Arguably the simplest possible daemon would be the one with semi-linear coupling to a single fast oscillator, as in Eqn.~(\ref{HOengine}) above. In rotating wave approximation (valid for large $\Omega$) and ignoring the possibility of operating as a daemon brake at negative weight velocity, this model has the adiabatically approximated
\begin{eqnarray}
V_\mathrm{eff} = g\,q + \sqrt{J-p-\epsilon g t}\,\cos(q)
\end{eqnarray}
which is a tilted washboard potential in which the barrier height is steadily falling, allowing no trajectories to be captured as in Fig.~2. 

This single-oscillator daemon could perhaps be called a `daemon of the first kind', because it allows bound trajectories of $q$ in which the engine runs, and the weight rises until the daemon engine stalls; but it can never ignite spontaneously. The phase space area enclosed by the engine-running separatrix decreases monotonically with time, for all initial conditions, because the barrier heights in the tilted washboard are always decreasing. Hence this daemon can keep running for some time, if it is running already, but it cannot begin running dynamically, for under incompressible phase space flow, no new trajectories can enter the shrinking volume of steady downconversion orbits. Indeed it is easy to verify, by numerical solution of the equations of motion for this system, that autonomous ignition simply never occurs.

The problem is even more serious than it may seem, because although this daemon can keep running for some possibly long time if it is running initially, controlling a system's initial conditions is not actually a trivial task. Only theoreticians can simply impose initial conditions at whim; in reality, they must be prepared through some process which is itself a dynamical evolution under some Hamiltonian. Unless a daemon has been running since the Big Bang, preparing it in an initially running state really means that it must begin in a state with no downconversion ongoing, and be brought into a running orbit by some state-preparation dynamics. If this preparation dynamics does not achieve control over fast variables, then it will amount only to some revision of $H_\mathrm{eff}$ during the initial phase of state preparation, and Liouville's theorem will still apply to the slower evolution, constraining the range of initial states which can be prepared. 

One can for example add an external starter force to the single-oscillator daemon, to make $g$ temporarily negative and boost the weight's speed up to the critical velocity from below. This will indeed enable autonomous ignition, by making the bound-orbit volume grow with time during the interval of negative $g$. Once the starter force is removed, however, the adiabatic invariance of enclosed phase space area will inevitably make the daemon engine stall before extracting more energy from $U$ than was put in by the starter force. The requirement for growing separatrix volume in order to achieve useful downconversion is not to be circumvented trivially.

The requirement is by no means impossible to satisfy, however. Consider the daemon model which is illustrated in our main text's Figure 2, for which the fuel subsystem consists of two oscillators, both of whose frequencies are high, but whose frequency difference plays the role of the high frequency $\Omega$:
\begin{eqnarray}\label{2osc}
U &=& \frac{P_+^2 + P_-^2}{2} + \frac{(\omega+\Omega)^2}{2}Q_+^2 + \frac{\omega^2}{2}Q_-^2\nonumber\\
V &=& kQ_+Q_-\cos(z/\lambda)\;.
\end{eqnarray}
This $U$ has the $\tau,U,\vec{\alpha}$ Hamilton-Jacobi representation
\begin{eqnarray}\label{2oscAAV}
Q_+ &=& \frac{\sqrt{2(U-\omega A)}}{\sqrt{\Omega(\Omega+\omega)}}\cos[(\Omega+\omega)\tau+\alpha]\nonumber\\
P_+ &=& -\frac{\sqrt{2(U-\omega A)(\Omega+\omega)}}{\sqrt{\Omega}}\sin[(\Omega+\omega)\tau+\alpha]\nonumber\\
Q_- &=& \frac{\sqrt{2[(\Omega+\omega) A-U]}}{\sqrt{\Omega\omega}}\cos[\omega\tau+\alpha]\nonumber\\
P_- &=& -\frac{\sqrt{2[(\Omega+\omega) A-U]\omega}}{\sqrt{\Omega}}\sin[\omega\tau+\alpha]\;,
\end{eqnarray}
where $(\alpha,A)$ are conjugate slow variables $\vec{\alpha}$. In rotating wave approximation valid for large $\Omega$ and for weight speeds not close to $dz/dt = \pm (\Omega+2\omega)\lambda$ or $dz/dt = -\Omega\lambda$, this model has
\begin{eqnarray}
\bar{V} = k\frac{\sqrt{(U-\omega A)[\Omega A-(U-\omega A)]}}{2\Omega\sqrt{\omega(\Omega+\omega)}}\,\cos(\lambda^{-1}z-\Omega\tau)\;.
\end{eqnarray}
This implies that $A$, which is the sum of the action variables of the two oscillators, is a constant of the motion. For fixed $A$, the minimum value of $U$ is $\omega A$, and so we can shift $U \to U+\omega A$. The maximum value of thus-shifted $U$ for this $A$ is then $U_{\mathrm{max}}=\Omega A$. In dimensionless variables we will therefore have
\begin{eqnarray}
V_{\mathrm{eff}} =g\,q + \sqrt{U(U_{\mathrm{max}}-U)}\,\cos(q)\;.
\end{eqnarray}

As with the single-oscillator model considered above, for small $\epsilon$ the phase space area of bound orbits is in this case determined (up to small corrections) by the barrier height $\sqrt{U(U_{max}-U)}$. The separatrix area increases with $t$, and autonomous ignition is possible, whenever this barrier height \textit{decreases} with $U = J - p -\epsilon g t$. Hence the condition for spontaneous ignition to be possible is that
\begin{equation}\label{cond}
\frac{d}{dU} U(U_{\mathrm{max}}-U) < 0 \Rightarrow U > U_{\mathrm{max}}/2
\end{equation}
must hold at the time when the weight's speed falls to $dz/dt = 1/\delta$. Since for small $\epsilon$ $U$ will not change appreciably until the daemon engine ignites (if it does ignite), this condition (\ref{cond}) may simply be applied directly to $U_I$, the initial value of $U$. The daemon will not ignite autonomously if its initial fuel level is less than half the maximum (given $A$), but if its `tank' is more than half full ($U_{I}>U_{\mathrm{max}}/2)$, then the fraction of initial conditions leading to autonomous ignition grows, until in the neighborhood of $U_I= U_{\mathrm{max}}$ the igniting fraction can even reach unity (the barriers spring up quite suddenly in this regime).

These two examples also show how the consequences of the volume-increase law for daemon ignition can be definite and non-trivial. As noted in our main text, $p = \epsilon/\delta + \mathcal{O}(\epsilon\delta^{0})$ on the separatrix, and so the behavior of $S(t)$ for fixed $J$ is given to leading order in $\epsilon$ by $S(U)$ (the phase space volume of bound orbits under $H_{\mathrm{eff}}$ with $V(q,J-p-\epsilon g t) \to V(q,U)$). This $S(U)$ can be evaluated by geometry and integration, without having to determine any actual motion from any particular initial conditions. Indeed for the whole class of effective Hamiltonians of the form
\begin{equation}\label{fqclass}
H_{\mathrm{eff}} = \frac{(p-\epsilon/\delta)^{2}}{2\epsilon}+\epsilon g q + \epsilon f(U) \cos(q)\;,
\end{equation}
we can determine the local maximum of $V_{\mathrm{eff}}$ to lie at $q\to q_{0}=\sin^{-1}(g/f)$ and hence find the separatrix energy $E_{0} = \epsilon g [q_{0}+\sqrt{(f/g)^{2}-1}]$. (The separatrix ceases to exist, because there are no bound orbits, for $|f|<|g|$.) This provides the general result
\begin{eqnarray}\label{SigmaX}
S(U) &=& \int p(E_{0},q)\,dq\nonumber\\
&=& 2 \int_{q_{0}}^{q_{1}}dq\,\sqrt{2\epsilon(E_{0}-V_{\mathrm{eff}}(q)}\nonumber\\
&=& \epsilon\sqrt{g}\tilde{S}\left(\frac{f(U)}{g}\right)\nonumber\\
\tilde{S}(X) &\equiv& 2\int_{q_{0}(X)}^{q_{1}(X)}\! dq\,\sqrt{q_{0}(X)+\sqrt{X^{2}-1}-q-X\cos(q)}\;,\end{eqnarray}
where $q_{2}(X)$ is the turning point of the highest bound orbit in $V_{\mathrm{eff}}$, \textit{i.e.} the second zero of $E_{0}-V_{\mathrm{eff}}$. The function $\tilde{S}(X)$ cannot be given in terms of elementary functions, but it is easily computed numerically. It has the limits $\tilde{S}(1)=0$ and $\tilde{S}(X)\to 8\sqrt{2X}$ for $X\to\infty$. 

Inserting $f(U)\to \sqrt{U}$ for the single-oscillator daemon model of (\ref{HOengine}) above, we find $S(U)$ as shown in Fig.~S-2(a). Since $U$ grows monotonically with $S$, no autonomous ignition is possible. Inserting instead $f(U)\to \sqrt{U(U_{\mathrm{max}}-U)}$ for the two-oscillator daemon model of (\ref{2osc}), we find the $U$-versus-$S$ relation shown in Fig.~S-2(b). The curve plotted in Fig.~S-2(b) has vertical mirror symmetry through the horizontal line $U=U_{\mathrm{max}}/2$. What Fig.~S-2(b) shows, therefore, is that the two-oscillator daemon can only ignite autonomously if its initial fuel level is above half of $U_{\mathrm{max}}$, because only for $U_{I}>U_{\mathrm{max}}/2$ will $U$ be decreasing with $S$, and hence $S$ be increasing with $t$, around the instant when the weight decelerates to $\dot{z}=1/\delta$. 

What Fig.~S-2(b) further shows is that, if the daemon does ignite autonomously, then it will stall at $U_{F}=U_{\mathrm{max}}-U_{I}$, so that the total work done is $2U_{I}-U_{\mathrm{max}}$. The reason for this is that, while the daemon engine runs, the area enclosed by its orbit remains adiabatically invariant, in accordance with the adiabatic theorem. If the daemon did ignite autonomously, it did so by entering the instantaneously outermost bound orbit, just inside the separatrix. Hence the area initially enclosed by its orbit was $S(U_{I})$, and so the adiabatically invariant enclosed area remains close to $S(U_{I})$. The daemon engine will therefore stall when downconversion has drained $U$ to the point $U=U_{F}$ at which $S(U_{F})=S(U_{I})$.  For daemons in general, the initial value $U_{I}$ thus determines $U_{F}$. In the present case, the horizontal mirror symmetry of $S(U)$ implies $U_{F}=U_{\mathrm{max}}-U_{I}$.

As shown in Fig.~S-3, numerically solving the system's full equations of motion for a range of initial conditions confirms these results robustly, up to small post-adiabatic corrections (\emph{i.e.} variations in $U_{F}$ at order $\epsilon$, which depend on the exact initial conditions). See Fig.~S-3.

\begin{figure*}[htbp]\label{figS2}
	\centering
	\subfigure[]{\includegraphics[width=.45\textwidth]{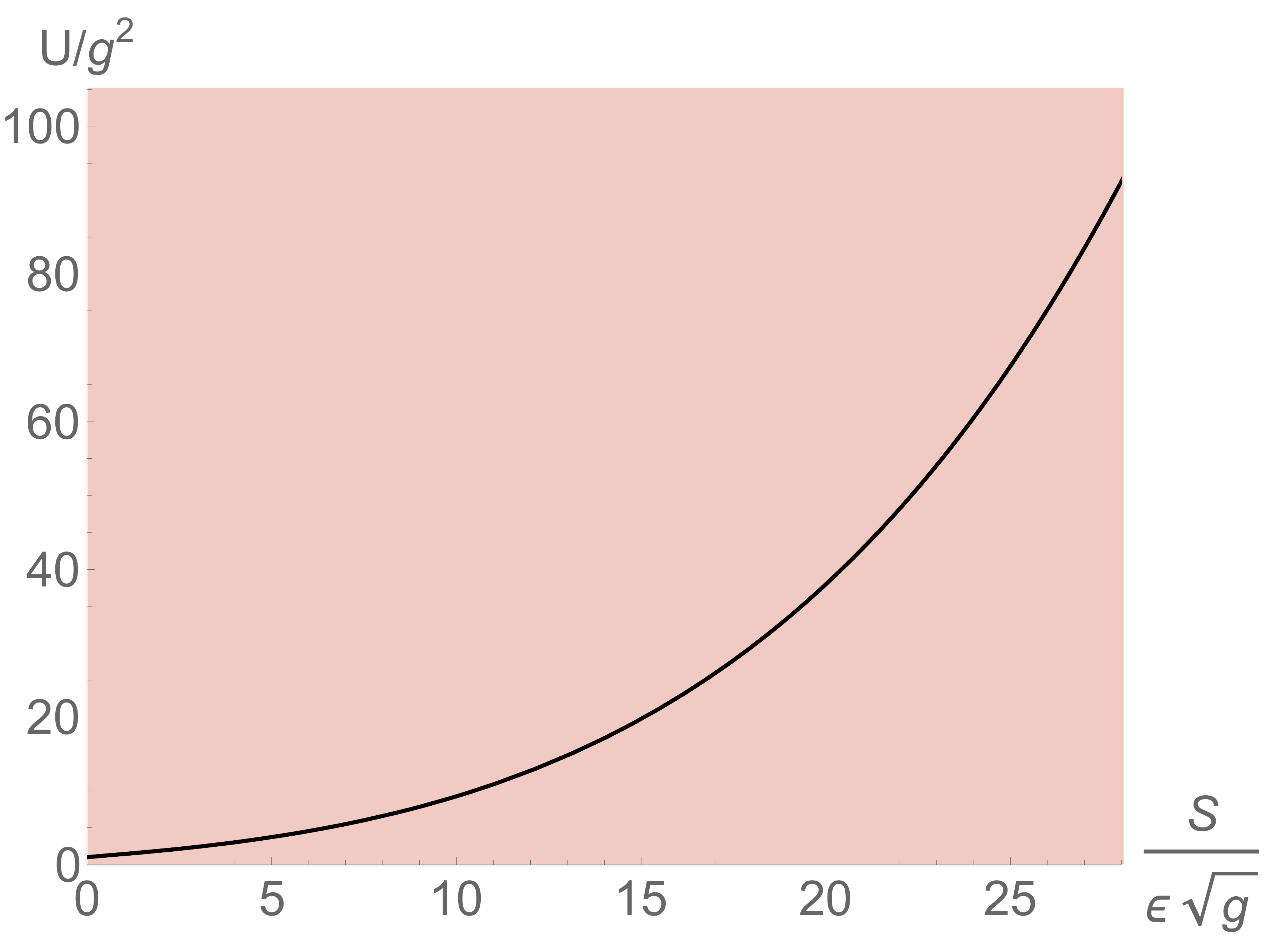}}
	\subfigure[]{\includegraphics[width=.45\textwidth]{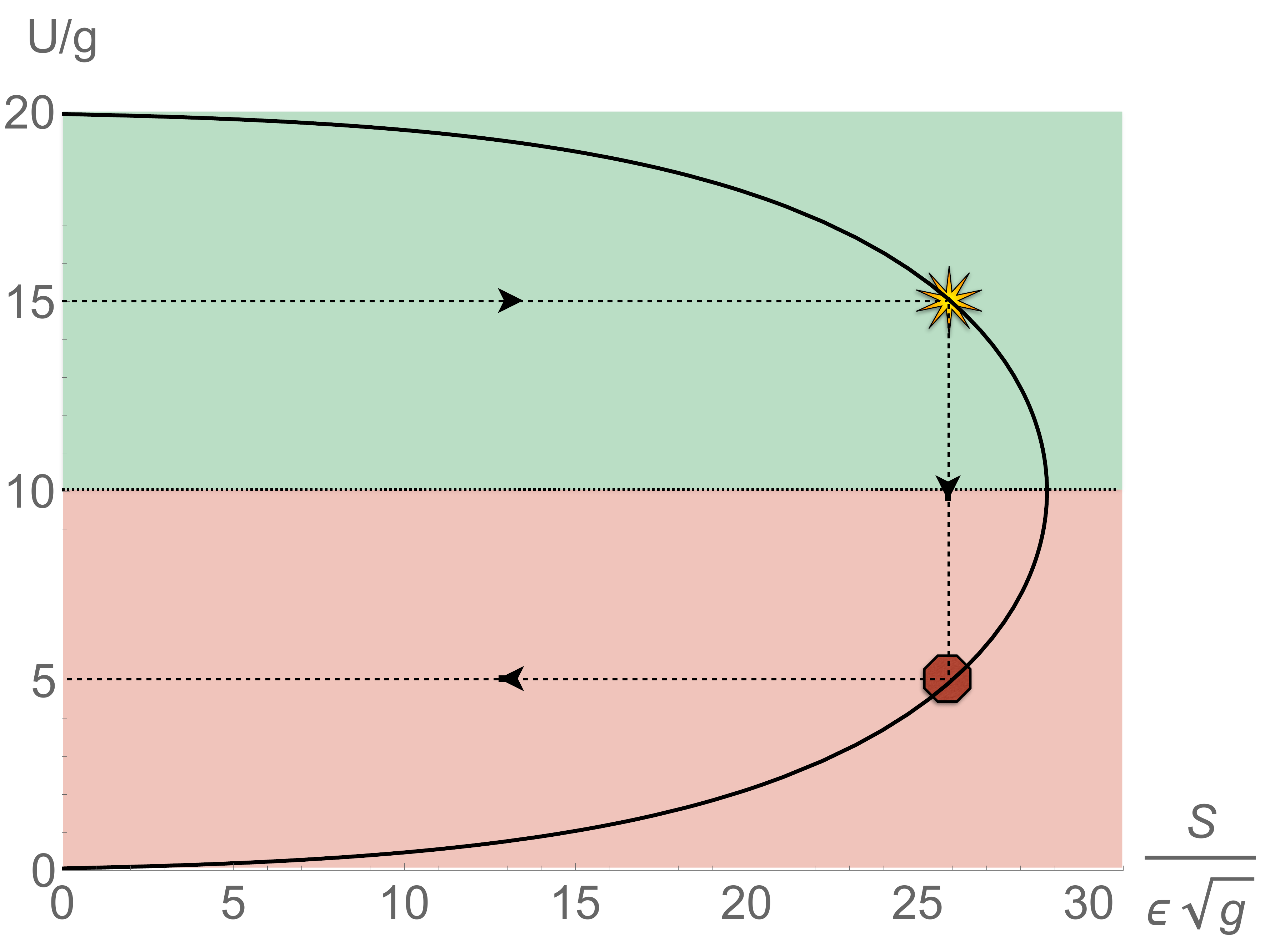}}
\caption{Curves of $U$ versus $S$ for (a) the single-oscillator daemon of Eqn.~(\ref{HOengine}), and (b) the two-oscillator daemon of Eqn.~(\ref{2osc}), for $U_{\mathrm{max}}=20g$. Compare with Fig.~3 of the main text. Red backgrounds indicate initial fuel energies $U_{I}$ from which spontaneous ignition is impossible, because the curve of $U$ versus $S$ has positive slope; green denotes energies at which the slope is negative, so ignition  can occur. The single-oscillator daemon can never ignite; it can perform steady downconversion, but only if this is specified as an initial condition. The two-oscillator daemon can ignite if its fuel tank is initially more than half full; if it does ignite, it will stall at a predictable final fuel level $U_{F}\doteq U_{\mathrm{max}}-U_{I}$, being the point at which the separatrix volume $S(U_{F})=S(U_{I})$. A daemon of this type can only convert all its fuel into work if it is initially fully charged ($U_{I}=U_{\mathrm{max}}$).}
\end{figure*}

The two-oscillator daemon thus obeys a non-trivial but manageable constraint. If its fuel tank is initially full, then all the fuel can be consumed; if it is initially 3/4-full, then the daemon engine will stall with 1/4 of its maximum fuel remaining; and so on. The daemon will not perform work at all if the fuel tank is initially less than half full. This limitation is not obvious from energetic considerations alone, but it follows from the constraints on phase space volumes (the adiabatic theorem, on one hand, and our main text's eqn.~(6), on the other). Having to fill the daemon fuel tank more than half-full before every re-use is a limitation which cannot be ignored, but if a useful device based on this model system were ever built, the requirement to recharge it more than half-full before every re-use should not be too severe for practical application.

\begin{figure*}[htbp]\label{figS3}
	\centering
	\subfigure[]{\includegraphics[width=.45\textwidth]{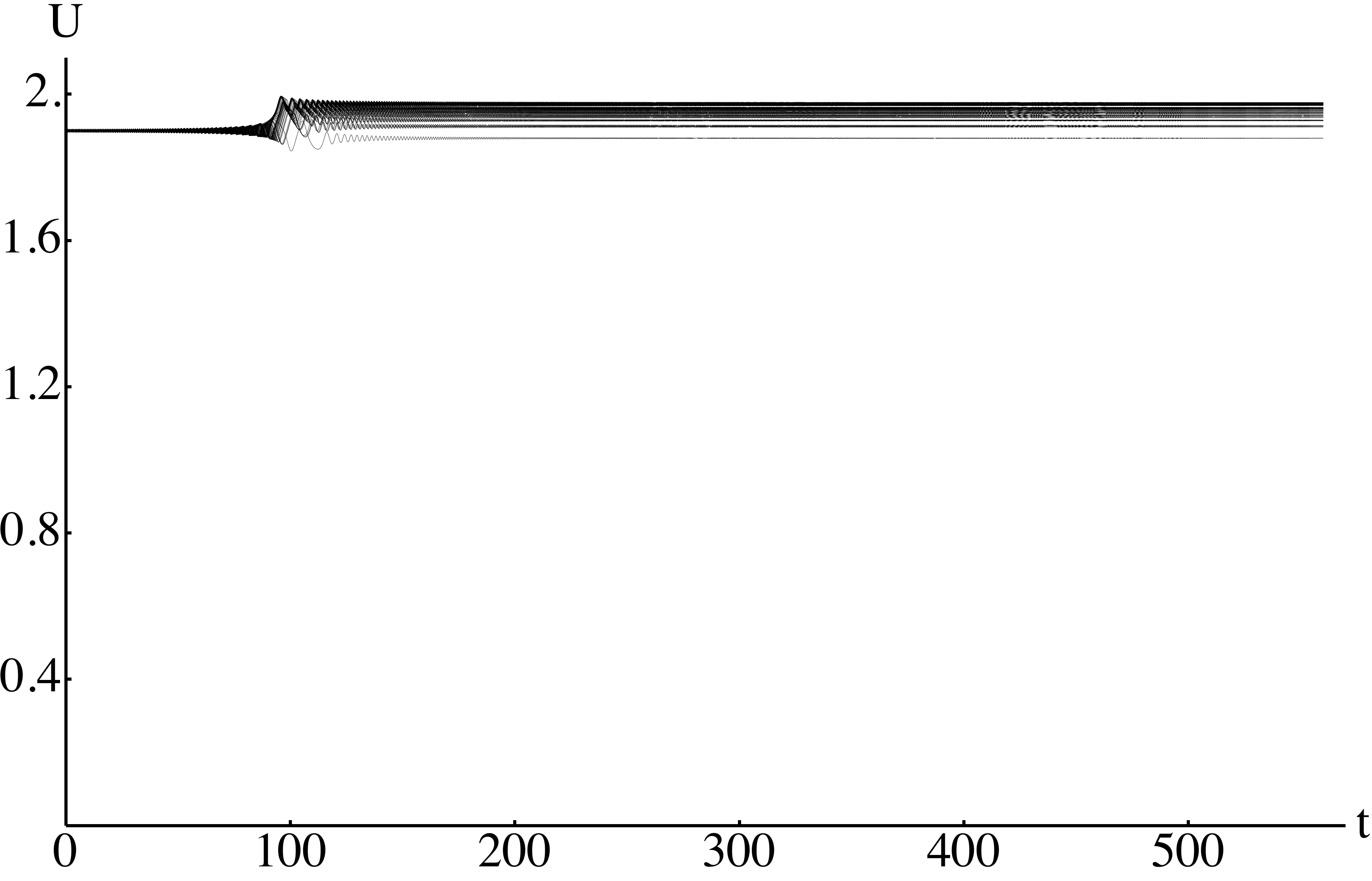}}
	\subfigure[]{\includegraphics[width=.45\textwidth]{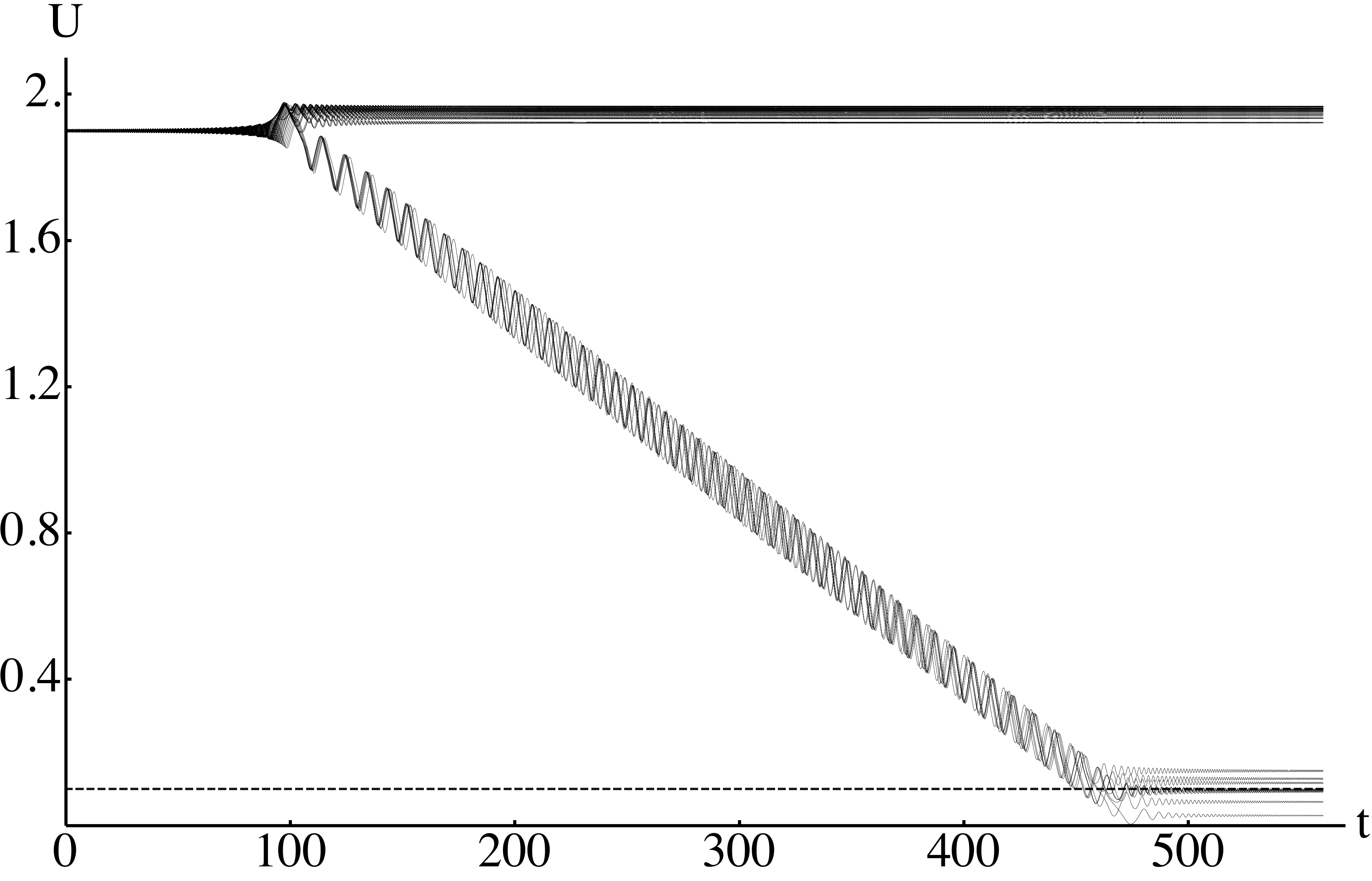}}
\caption{Fuel energy $U$ versus time $t$, as given by numerical solutions to full equations of motion for daemon systems with (a) a single-oscillator fuel source similar to Eqn.~(\ref{HOengine}), and (b) the two-oscillator fuel of (\ref{2osc}). All trajectories shown have $\epsilon=0.05$, $\delta=0.1$, $g=0.1$, $p(0)= 2\epsilon/\delta = 1$, $z(0)=0$ and $U(0)=1.9$. In a) the coupling is $\epsilon V=-\epsilon \sqrt{0.1 U}\cos(z-\tau)$ while in b) it is $\epsilon V = -\epsilon \sqrt{(2-U)U}\cos(z-\tau)$, so that the effective potentials for $q=z-\tau$ are initially identical in the two cases. Each plot shows a collection of trajectories with 50 different values of $\tau(0)$, evenly distributed in $[0,2\pi]$. In all cases the decelerating weight hits resonant speed at around $t=100$.  As a) shows, the single-oscillator daemon never ignites, whereas for the two-oscillator daemon, a finite fraction of initial conditions (9 of the 50 trajectories) are seen in b) to lead to spontaneous ignition. Since initial fuel energies $U_{I}=1.9$ are equal for all initial conditions represented, the adiabatic theory of the main text predicts equal final fuel $U_{F}=U_{I}$ for all non-igniting trajectories, and also equal final fuel ($S(U_{F})=S(U_{I})\Rightarrow U_{F}=2-U_{I}=0.1$) for all igniting trajectories in b). In fact there is a finitely narrow range of final $U$ in both cases --- a post-adiabatic effect of order $\epsilon$. Note that for non-igniting trajectories $U_{F}$ is almost always slightly \emph{greater} than $U_{I}$.  This small rise in $U$ results from the detailed interplay of forces on $z$ and $\tau$ along particular trajectories, but it can also be deduced from incompressible flow in the full system's phase space (including $U$): engine-running orbits exist for both types of daemon engine (though in the single-oscillator case they can only be attained as initial conditions), and these represent a certain volume of phase space that is steadily migrating downward from high $U$ to low. To make room for this small volume of steadily descending phase space, non-running orbits are all displaced slightly upward in $U$, in a kind of phase space convection.}
\end{figure*}

\clearpage
\section{Appendix: Multiple time scales}

Although resonance is our reason for letting $\tilde{V}\to\bar{V}$, our resonance occurs only at a special speed of the weight, so this is not like the resonance of two linearly coupled harmonic oscillators, which obtains either throughout all of phase space at once, or not at all. Moreover, resonance in linear systems means frequency matching, but the weight's own natural motion, if $\epsilon V\to0$, is to accelerate steadily downwards, rather than oscillate at any frequency. We must therefore speak more generally and technically about what we mean by resonance in our case.

The generic daemon Hamiltonian possesses two small parameters, which imply a double hierarchy of time scales. Dynamics involving multiple time scales is the default in nature, and dealing with them is essential to physics. The discoveries of the twentieth century revealed the universe as an extremely large system composed of extremely tiny moving parts. The vast range of length scales involves a correspondingly vast range of time scales. Even ordinary daily life depends on processes with time scales ranging from femtoseconds to hours.

Evolution on multiple time scales is determined by equations of motion that involve small dimensionless parameters. This elementary fact implies an inherent subtlety, however, in distinguishing changes that are small from changes that are slow. A differential equation of the form
\begin{equation}
\frac{d f}{dt} = \mathcal{O}(\epsilon)\;,
\end{equation}
which sets the time derivative of $f(t)$ equal to some quantity of order $\epsilon\ll 1$, is often assumed to have a solution of the form $f =\epsilon f_\mathrm{small}(t)$. Depending on exactly what the small quantity on the right side of the equation happened to be, however, the solution might instead be of the form $f = f_\mathrm{slow}(\epsilon t)$. In the former case $f(t)$ is small, at least for some time, while in the latter case it is slow, but not necessarily small at all.

Which of these scenarios will occur may not always be immediately obvious by inspection, but the difference between them is easy to recognize in solutions. In a case such as
\begin{equation}
\frac{d f}{dt} = \epsilon g(t) + \epsilon h(\epsilon t)\;,
\end{equation}
where $g$ and $h$ are bounded and their dependences on their respective arguments do not involve any large or small parameters, one clearly finds $f(t) = \epsilon f_1(t) + f_2(\epsilon t)$ where $\dot{f}_{1}(t)=g(t)$ and $\dot{f}_{2}(t)=h(t)$. If one were willing to ignore quantities of order $\epsilon$ as insignificantly small, then one could solve for $f$ with sufficient accuracy by setting $\epsilon g(t)\to 0$; but if one needed to know the behavior of $f$ for $t$ as large as $\mathcal{O}(\epsilon^{-1})$, then one would need to retain $\epsilon h(\epsilon t)$, even though it is small.

This subtlety persists quite generally in differential equations, and is not always so easy to resolve. In such a more general case as
\begin{equation}
\frac{d f}{dt} = g(f,t)+\mathcal{O}(\epsilon)\;,
\end{equation}
whether the effects of the small $\mathcal{O}(\epsilon)$ perturbation are small or slow depends on the precise forms of both $g(f,t)$ and the perturbation. To approximate accurately over long time scales in non-trivial cases, one must keep track of slowness and smallness separately, while solving non-trivial differential equations.

A formal method for doing this is \textit{multiple scale analysis}. It begins with the apparently complicating step of embedding the single dimension of physical time in a space of more time dimensions. In our three-scale case, we introduce three distinct time axes, 
\begin{equation}
t\to \vec{t}=(t_1,t_2,t_3)\;.
\end{equation}
For each time-dependent dynamical variable $X(t)$ (meaning any of the $q_n(t)$ or $p_n(t)$), we then seek a function of all three time variables, $X(\vec{t})$. The idea in doing this is that the fast evolution of $X(t)$ should be represented through the dependence of $X(t_1,t_2,t_3)$ on $t_1$, the intermediate time scale evolution of $X(t)$ by the dependence of $X(\vec{t})$ on $t_2$, and the slow evolution of $X(t)$ by the dependence of $X(\vec{t})$ on $t_3$. That is, $t_{1}$ is the `fast time' (it will be compressed by the factor $1/\delta$ in physical time dependence), $t_{2}$ is the `intermediate time', and $t_{3}$ is the `slow time' (it will be stretched out by a factor $1/\epsilon$ in physical time dependence).

This intention of representing multiple time scales with multiple times is made concrete by re-interpreting
\begin{equation}\label{treplace}
\frac{d}{dt}\longrightarrow \frac{1}{\delta}\frac{\partial}{\partial t_1} + \frac{\partial}{\partial t_2} + \epsilon\frac{\partial}{\partial t_3}
\end{equation}
in the equations of motion (main text Eqn.~(1)), so that $X(t/\delta, t, \epsilon t)$ provides a solution to the original equations of motion for $X(t)$. The behavior of $X(\vec{t})$ along the line $\vec{t} = (t/\delta,t,\epsilon t)$ thus reproduces the physical time evolution of $X(t)$, and the dependence of $X(\vec{t})$ on its three time arguments is directly mapped into dependence of $X(t)$ on the single time $t$, but with three different scale factors. In our case, for example, the Hamiltonian equations of motion (1) (main text) with $\tilde{H}$ from (\ref{tildeH}) (with $\tilde\ $ accents dropped) read 
\begin{eqnarray}\label{MSAeqs}
 \left[\frac{1}{\delta}\frac{\partial}{\partial t_1} + \frac{\partial}{\partial t_2} + \epsilon\frac{\partial}{\partial t_3}\right]\tau &=& \frac{1}{\delta} + \epsilon\frac{\partial}{\partial U}V\nonumber\\
 \left[\frac{1}{\delta}\frac{\partial}{\partial t_1} + \frac{\partial}{\partial t_2} + \epsilon\frac{\partial}{\partial t_3}\right]U &=& - \epsilon\frac{\partial}{\partial \tau}V\nonumber\\
 \left[\frac{1}{\delta}\frac{\partial}{\partial t_1} + \frac{\partial}{\partial t_2} + \epsilon\frac{\partial}{\partial t_3}\right]^2 z &=& -g -\frac{\partial}{\partial z}V\;.
\end{eqnarray}

Since small parameters appear in these equations, we can approximate perturbatively by expanding each variable as a power series in $\delta$ and $\epsilon$:
\begin{equation}
X = \sum_{m,n} \delta^m \epsilon^n X_{mn}(\vec{t})\;.
\end{equation}
Now we can explain what has been gained by extending time to three dimensions. Away from the physical time line $(t/\delta,t,\epsilon t)$, the behavior of $X(\vec{t})$ on unphysical $\vec{t}$ may be chosen \emph{arbitrarily}. The strategy of multiple scale analysis is to use this freedom judiciously, to separate slow change from small by ensuring that no $X_{mn}$ ever grow large. Effects identified as small thus remain small forever, while slow processes are treated accurately as such.

To show how this works in general, and at the same time derive the specific $z-\tau$ dependence in our main text's daemon Hamiltonian (2), it suffices to examine evolution of some low-order $z_{mn}$, $\tau_{mn}$, and $U_{mn}$. (The $\vec{\alpha}$ will not be involved in this discussion because the equations up to order $\epsilon$ only confirm that $\vec{\alpha}_{00}$ can depend only on $t_3$). 

In (\ref{MSAeqs}) the single $\delta^{-2}\epsilon^0$ term yields 
\begin{equation}
\frac{\partial^2}{\partial t_1^2}{z_{00}} = 0\;,
\end{equation}
which implies
\begin{equation}
z_{00} = v_0(t_2,t_3)t_1 + \bar{z}(t_2,t_3)\;.
\end{equation}
The $\delta^{-1}\epsilon^0$ equations read
\begin{eqnarray}
\frac{\partial}{\partial t_1}\tau_{00} &=& 1\nonumber\\
\frac{\partial}{\partial t_1}U_{00} &=& 0\nonumber\\
\frac{\partial^2}{\partial t_1^2}{z_{10}} + 2\frac{\partial^2}{\partial t_1\partial t_2}z_{00} &=&0\;,
\end{eqnarray}
which yield
\begin{eqnarray}
\tau_{00} &=& t_1 + \bar{\tau}_{00}(t_2,t_3)\nonumber\\
U_{00} &=& \bar{U}_{00}(t_2,t_3)\nonumber\\
z_{10} &=& -\frac{\partial v_0}{\partial t_2}t_1^2+v_1(t_2,t_3) t_1+\bar{z}_{10}(t_2,t_3).
\end{eqnarray}
But now we note that on the physical time axis the term $t_1v_0(t_2,t_3)$ and the term $ t_1^2\delta\,\partial_{t_2}v_0(t_2,t_3)$ both represent change on the same $t/\delta$ scale, and that for large $t$ the formally smaller second term will actually be larger than the first term. We would prefer to have a perturbation series that converged for all $t$; and \textit{we are free to have our wish}. The behavior of $v_0$ as a function of $t_2$ and $t_3$ has not been constrained by our equations of motion, and our equations of motion express the entire physically required behavior in the $\vec{t}$ space. We are thus free to impose any conditions we please on $v_0(t_2,t_3)$. By choosing $v_0 = v_0(t_3)$, we can obtain $z_{10} = v_1(t_2,t_3) t_1 + \bar{z}_{10}(t_2,t_3)$, which does not compete in size with $z_{00}$ for any value of $t$. 

In other words, the behavior of $v_0$ as a function of $t_2$ has not been fixed by the equation of motion directly, but has in effect been determined by our goal of solving the equations of motion with a perturbation series that still converges for large $t$. This is the essential mechanism of multiple scale analysis.

Finally at order $\delta^0\epsilon^0$ we find something less trivial:
\begin{eqnarray}
\frac{\partial}{\partial t_1}\tau_{10} + \frac{\partial}{\partial t_2}\tau_{00} &=& 0\nonumber\\
\frac{\partial}{\partial t_1}U_{10} + \frac{\partial}{\partial t_2}U_{00}  &=& 0\nonumber\\
\frac{\partial^2}{\partial t_1^2}{z_{20}}+2\frac{\partial^2}{\partial t_1\partial t_2}z_{10}+\frac{\partial^2}{\partial t_2^2}z_{00} &=& - g - \frac{\partial}{\partial z_{00}}V\left(z_{00},\tau_{00},U_{00},\vec{\alpha}_{00}\right)\;.
\end{eqnarray}
Similar considerations to those immediately above now lead us to conclude $\bar{\tau}_{00} = \bar{\tau}_{00}(t_3)$, $\bar{U}_{00} = \bar{U}_{00}(t_3)$, and $v_1 = v_1(t_3)$. 

We are left with a genuinely non-trivial equation
\begin{eqnarray}\label{GNT}
\frac{\partial^2}{\partial t_1^2}{z_{20}}+\frac{\partial^2}{\partial t_2^2}\bar{z}_{00} &=& - g - \frac{\partial}{\partial \bar{z}_{00}}V\left(v_0(t_3)t_1+\bar{z}_{00},t_1+\bar{\tau}_{00}(t_3),\bar{U}_{00}(t_3),\vec{\alpha}_{00}(t_3)\right)\;.
\end{eqnarray}
Since $V$ is a bounded function whose dependence on $\tau$ involves no small parameters, any terms on the right side of (\ref{GNT}) that depend on $t_1$ will merely imply some bounded $z_{20}$. Since $z_{20}$ contributes to $z$ at order $\delta^2$, this represents only very small, rapid oscillations or fluctuations in the height of the weight. Such small effects are of no concern in this paper. But what if there is some part of the $V$ (call it $\bar{V}$) on the right side of (\ref{GNT}) that does \textit{not} depend on $t_1$, but only on $t_2$ (through $z_{00}$) and $t_3$?  These would lead to a term $\sim t_1^2$ in $z_{20}$, breaking the convergence of our perturbation series at large $t$ --- unless we \textit{choose}
\begin{eqnarray}\label{GNTslow}
\frac{\partial^2}{\partial t_2^2}\bar{z}_{00} &=& - g - \frac{\partial}{\partial \bar{z}_{00}}\bar{V}\;.
\end{eqnarray}

Since our intent to describe microscopic engines rather than clockwork requires $V$ to depend (without small parameters) on $\tau$, it may well be that no terms in $V$ are independent of $t_1$, and so $\bar{V}=0$. (For $V$ to include some term that did not depend on $\tau$ at all would effectively describe a clockwork system, rather than one that could perform downconversion.) Having $\bar{V}=0$ would mean that the small, fast variations represented by $z_{20}$ were the only effect of the $\epsilon V$ coupling, and that the weight essentially accelerated downward under gravity just as if the daemon were not there at all. Other than to capture the tiny fluctuations, the whole multiple scale analysis would have been unnecessary, because as far as large scale, long term evolution is concerned, we could simply have replaced $V\to 0$ from the beginning.

In a system with $V\to0$, $U$ and $p+\epsilon g t$ would each separately be exactly constant. In a system with $V\not=0$, but with $\bar{V}=0$, $U$ and $p+\epsilon g t$ would not be exactly constant; but their only change would be small, rapid variations that would never accumulate into any significant alteration. One refers to quantities that are practically but not strictly constant in this sense as \textit{adiabatic invariants}, and to the scenario where the weight and the fuel are practically though not strictly decoupled as \textit{adiabatic decoupling}.

Adiabatic decoupling is the default expectation for weakly interacting degrees of freedom with disparate time scales. It expresses in formal terms the conclusion familiar from school physics, that significant energy exchange between weakly coupled systems is possible only when their frequencies are \textit{in resonance} --- that is, closely matching. 

There is, however, one way for the daemon's fuel not to decouple adiabatically from the weight. That is for $V$ to include some term of the form $V(z-v_{c} \tau, U,\vec{\alpha})$ for some constant $v_c$. (As emphasized above, it is \textit{not} necessary for $V$ to be entirely of this form.) In any such case, $\bar{V}$ will be nonzero whenever $v_0 = v_c$. If this $\bar{V}$ merely includes local minima deep enough to overcome gravity's $-g$, then the weight may steadily rise at physical speed $v_c \lambda/\delta$. Since the length scale $\lambda$ is microscopic, this is not necessarily a high speed in physical units, but its relatively large size $\sim 1/\delta$ achieves the daemon's purpose of extracting a large amount of steady work from high-frequency energy. This is still not a simple case of clockwork, however, with energy transfer among degrees of freedom with matching time scales. The resonance in this case is highly nonlinear, occurring only when the weight is rising at the particular critical speed $v_c$. And it leads not to oscillation of energy back and forth between weight and fuel, but to steady rise of the weight as fuel drains.

We identify $v_c/\delta$ as the steady speed at which the daemon is designed to lift the weight. Since the length scale $\lambda$ is in principle arbitrary, we can choose it without loss of generality to make $v_c = 1$. 

If two otherwise different $V$ should have the same $\bar{V}$ component, having the same dependence on $z-\tau$, then their respective $z_{20}$ terms might differ, but they would yield the same evolution for $z_{00}$. In particular they would both have the same $z_{00}$ evolution that would be found by replacing $V\to \bar{V}(z-\tau,U,\vec{\alpha})$ from the beginning. This is what we mean in our main text by saying that a daemon can be expressed with the particular form of $H$ given in our main text's Eqn.~(2), after discarding small, non-resonant terms because their effects are negligible.

\end{document}